\documentclass[twocolumn]{aastex631}
\usepackage[T1]{fontenc}
\usepackage[utf8]{inputenc}
\received{September 22, 2025}
\revised{November 25, 2025}
\accepted{December 3, 2025}
\shorttitle{\texttt{EzTaoX}: Scalable \& Robust AGN Light Curve Modeling}
\graphicspath{{./}{figs/}}
\usepackage{amsmath}
\usepackage{CJK}

\newcommand{\lledd}{$L/L_{\rm Edd}$}

\newcommand{\lbol}{$L_{\rm bol}$}

\newcommand{\msun}{$M_{\odot}$}

\newcommand{\sigmadrw}{$\sigma_{\mathrm{DRW}}$}

\newcommand{\mbh}{$M_{\rm {BH}}$}

\newcommand{\pkg}{\texttt{EzTaoX}}
\newcommand{\jl}{\texttt{JAVELIN}}
\newcommand{\tp}{\texttt{tinygp}}
\newcommand{\taudrw}{$\tau_{\rm DRW}$}
\renewcommand{\edit}[1]{{#1}}

\begin{document}
\begin{CJK*}{UTF8}{gbsn}
\title{Scalable and Robust Multiband Modeling of AGN Light Curves in Rubin-LSST}
\correspondingauthor{Weixiang Yu}
\email{wyu@ubishops.ca}

\author[0000-0003-1262-2897]{Weixiang Yu~(于伟翔)}
\affil{Department of Physics \& Astronomy, Bishop's University, 2600 rue College, Sherbrooke, QC, J1M 1Z7, Canada}

\author[0000-0001-8665-5523]{John~J.~Ruan}
\affil{Department of Physics \& Astronomy, Bishop's University, 2600 rue College, Sherbrooke, QC, J1M 1Z7, Canada}

\author[0000-0001-9947-6911]{Colin~J.~Burke}
\affil{Department of Astronomy, Yale University, 266 Whitney Avenue, New Haven, CT 06511, USA}

\author[0000-0002-9508-3667]{Roberto~J.~Assef}
\affil{Instituto de Estudios Astrof\'isicos, Facultad de Ingenier\'iay Ciencias, Universidad Diego Portales, Av. Ej\'ercito Libertador 441, Santiago, Chile}

\author[0000-0001-8211-3807]{Tonima T. Ananna}
\affil{Department of Physics and Astronomy, Wayne State University, 666 W. Hancock St, Detroit, MI, 48201, USA}

\author[0000-0002-8686-8737]{Franz~E.~Bauer}
\affil{Instituto de Alta Investigaci{\'{o}}n, Universidad de Tarapac{\'{a}}, Casilla 7D, Arica, Chile}

\author[0000-0001-7208-5101]{Demetra De Cicco}
\affil{Department of Physics, University of Napoli ``Federico II", via Cinthia 9, 80126 Napoli, Italy}
\affil{INAF - Osservatorio Astronomico di Capodimonte, via Moiariello 16, 80131 Napoli, Italy}
\affil{Millennium Institute of Astrophysics (MAS), Nuncio Monse\~nor S\'otero Sanz 100, Providencia, Santiago, Chile}

\author[0000-0003-1728-0304]{Keith Horne}
\affil{SUPA School of Physics and Astronomy, North Haugh, St. Andrews, KY16 9SS, Scotland, UK}

\author[0000-0002-8606-6961]{Lorena Hern\'{a}ndez-Garc\'{i}a}
\affil{Instituto de Estudios Astrof\'isicos, Facultad de Ingenier\'iay Ciencias, Universidad Diego Portales, Av. Ej\'ercito Libertador 441, Santiago, Chile}
\affil{Centro Interdisciplinario de Data Science, Facultad de Ingenier\'ia y Ciencias, Universidad Diego Portales, Av. Ej\'ercito Libertador 441, Santiago, Chile}
\affil{Millennium Institute of Astrophysics (MAS), Nuncio Monse\~nor S\'otero Sanz 100, Providencia, Santiago, Chile}
\affil{Millennium Nucleus on Transversal Research and Technology to Explore Supermassive Black Holes (TITANS), Gran Breta\~na 1111, Playa Ancha, Valpara\'iso, Chile}

\author[0000-0002-1134-4015]{Dragana Ili\'c}
\affil{Department of Astronomy, Faculty of Mathematics, University of Belgrade, Studentski trg 16, 11000 Belgrade, Serbia}
\affil{Hamburger Sternwarte, Universitat Hamburg, Gojenbergsweg 112, D-21029 Hamburg, Germany}

\author[0000-0002-3277-6335]{Vivek Kumar Jha}
\affil{National Centre for Radio Astrophysics - Tata Institute of Fundamental Research (NCRA-TIFR), Pune, India}

\author[0000-0001-5139-1978]{Andjelka B. Kova\v cevi\'c}
\affil{Department of Astronomy, Faculty of Mathematics, University of Belgrade, Studentski trg 16, 11000 Belgrade, Serbia}

\author[0000-0002-1380-1785]{Marcin Marculewicz}
\affil{Department of Physics and Astronomy, Wayne State University, 666 W. Hancock St, Detroit, MI, 48201, USA}

\author[0000-0002-5854-7426]{Swayamtrupta Panda}
\affil{International Gemini Observatory/NSF NOIRLab, Casilla 603, La Serena, Chile}

\author[0000-0001-5231-2645]{Claudio Ricci}
\affil{Instituto de Estudios Astrof\'isicos, Facultad de Ingenier\'iay Ciencias, Universidad Diego Portales, Av. Ej\'ercito Libertador 441, Santiago, Chile}
\affil{Department of Astronomy, University of Geneva, ch. d’Ecogia 16, 1290, Versoix, Switzerland}

\author[0000-0002-1061-1804]{Gordon T. Richards}
\affiliation{Department of Physics, Drexel University, 32 S.\ 32nd Street, Philadelphia, PA 19104, USA}

\author[0000-0003-0483-3723]{Rogemar A. Riffel}
\affil{Departamento de F\'{i}sica, CCNE, Universidade Federal de Santa Maria, 97105-900 Santa Maria, RS,Brazil}
\affil{Laborat\'orio Interinstitucional de e-Astronomia - LIneA, Rua Gal. Jos\'e Cristino 77, Rio de Janeiro, RJ - 20921-400, Brazil}

\author[0000-0001-7240-7449]{Donald P. Schneider}
\affil{Department of Astronomy \& Astrophysics, The Pennsylvania State University, University Park, PA 16802, USA}
\affil{The Institute for Gravitation for and the Cosmos, The Pennsylvania State University, University Park, PA 16802, USA}

\author[0000-0003-0820-4692]{Paula S\'{a}nchez-S\'{a}ez}
\affil{European Southern Observatory, Karl-Schwarzschild-Strasse 2, 85748 Garching bei M\"unchen, Germany}

\author[0009-0003-0654-6805]{Sarath Satheesh-Sheeba}
\affil{Instituto de Astrof\'{i}sica, Facultad de Ciencias Exactas, Universidad Andr\'{e}s Bello, Fern\'andez Concha 700, 7591538 Las Condes, Santiago, Chile}

\author[0000-0002-6562-8654]{Francesco Tombesi}
\affil{Physics Department, Tor Vergata University of Rome, Via della Ricerca Scientifica 1, 00133 Rome, Italy}
\affil{INAF – Astronomical Observatory of Rome, Via Frascati 33, 00040 Monte Porzio Catone, Italy}
\affil{INFN – Rome Tor Vergata, Via della Ricerca Scientifica 1, 00133 Rome, Italy}

\author[0000-0001-8433-550X]{Matthew J. Temple}
\affil{Centre for Extragalactic Astronomy, Department of Physics, Durham University, South Road, Durham DH1 3LE, UK}

\author[0000-0001-7416-9800]{Michael S. Vogeley}
\affiliation{Department of Physics, Drexel University, 32 S.\ 32nd Street, Philadelphia, PA 19104, USA}

\author[0000-0001-9163-0064]{Ilsang Yoon}
\affil{National Radio Astronomy Observatory, 520 Edgemont Road, Charlottsville, VA 22904, USA}
\affil{Department of Astronomy, University of Virginia, 530 McCormick Rd, Charlottesville, VA 22904, USA}

\author[0000-0002-4436-6923]{Fan Zou}
\affil{Department of Astronomy, University of Michigan, 1085 S University, Ann Arbor, MI 48109, USA}

\begin{abstract}
The Vera C. Rubin Observatory's Legacy Survey of Space and Time (LSST) will monitor tens of millions of active galactic nuclei (AGNs) for a period of 10 years with an average cadence of 3 days in six broad photometric bands.
This unprecedented dataset will enable robust characterizations of AGN UV/optical variability across a wide range of AGN physical properties.
However, existing tools for modeling AGN light curves are not yet capable of fully leveraging the volume, cadence, and multiband nature of LSST data.
We present \pkg, a scalable light curve modeling tool designed to take advantage of LSST's multiband observations to simultaneously characterize AGN UV/optical stochastic variability and measure interband time delays.
\pkg~achieves a speed increase of $\sim10^2-10^4\times$ on CPUs over current tools with similar capabilities, while maintaining equal or better accuracy in recovering simulated variability properties. 
This performance gain enables continuum time-delay measurements for all AGNs discovered by LSST---both in the Wide Fast Deep survey and the Deep Drilling Fields---thereby opening new opportunities to probe AGN accretion-flow geometries. 
In addition, \pkg's multiband capability allows robust characterization of AGN stochastic variability down to hourly timescales, facilitating the identification of accreting low-mass AGNs---such as those residing in dwarf galaxies---through their distinctive variability signatures.
\end{abstract}

\section{Motivation}\label{sec:motivation}

The UV/optical continuum luminosity of active galactic nuclei (AGNs) exhibit correlated stochastic variability on timescales ranging from hours to years~\citep[e.g.,][]{vandenberk2004, sesar2007, schmidt2012}.
This variability is believed to originate in the accretion disk, potentially driven in part by thermal reprocessing of X-ray photons emitted from the central corona~\citep{Krolik1991} and/or accretion rate fluctuations propagating through the disk~\citep{lyubarskii1997}.
For a recent review of the observational properties of AGN variability in the UV/optical and X-ray bands, we refer the reader to \citet{Paolillo2025}.

A damped random walk (DRW) process is commonly adopted to model AGN UV/optical light curves~\citep{kelly2009}. The power spectral density (PSD) of a DRW exhibits a broken power-law (PL; $1/f^{\alpha}$) form, where the PL exponent $\alpha$ transitions from zero at low frequencies to two at high frequencies. The point at which this change in PL exponent occurs is known as the break frequency, which corresponds to a characteristic timescale denoted by $\tau_{\rm DRW}$. The asymptotic variability amplitude of the process is denoted by~\sigmadrw. 
Although the DRW process is a phenomenological model and the precise PSD shape of intrinsic AGN UV/optical variability remains under debate~\citep[e.g.,][]{mushotzky2011, kasliwal2015, smith2018, Arevalo2024, Beard2025, Petrecca2024, Yu2025}, it continues to be a valuable and widely used tool for modeling AGN light curves because of its adequacy and computational efficiency.

Modeling AGN UV/optical light curves as a DRW has revealed compelling correlations between variability and AGN fundamental properties.
Numerous previous investigations have shown that the best-fit~\sigmadrw~correlates negatively with the Eddington ratio (\lledd) and bolometric luminosity (\lbol) of AGNs~\citep[e.g.,][]{macleod2010, sanchez-saez2018, suberlak2021}.
Additionally, the characteristic timescale of the DRW model (\taudrw) correlates positively with \mbh~\citep[e.g.,][]{burke2021, Wang2023a}.
These empirical relations not only provide valuable insights into the physical mechanisms driving AGN variability, but also form the basis for new techniques to estimate AGN fundamental properties using photometric data alone. 
For example, \citet{burke2022} leveraged the $\tau_{\rm DRW}$---\mbh~correlation to search for low-mass AGNs powered by accreting massive black holes (MBH; $10^5\,$\msun$\,<\,$\mbh$\,<\,$$10^7\,$\msun) at the centers of dwarf galaxies
~\citep[i.e., $M_\star \lesssim 10^{10}\,M_\odot$;][]{greene2020}.

AGN UV/optical variability at longer wavelengths is also often observed to lag behind the variability at shorter wavelengths~\citep{Collier1999, sergeev2005}.
Under the thermal reprocessing scenario, longer-wavelength photons originate from disk annuli located farther from the central supermassive black hole (SMBH) than those emitting shorter-wavelength photons~\citep{shakura1973}. 
Consequently, the reprocessed emission at longer wavelengths exhibits longer light-crossing time delays relative to the central irradiating source~\citep{Cackett2007}.
This specific geometry---in which a compact source near the SMBH (e.g., an X-ray corona) irradiates the accretion disk---is commonly referred to as the lamp-post model.
In this framework, the size of the accretion disk can be inferred by measuring the time lags between UV/optical light curves at different wavelengths---a technique known as continuum reverberation mapping (CRM), which probes the geometry of unresolved sources through time-delay measurements~\citep[see][and references therein]{cackett2021}.

Numerous CRM experiments have reported that AGN accretion disk sizes are systematically larger than those predicted by the standard thin-disk model~\citep[e.g.,][]{fausnaugh2016, jiang2017, Jha2022}.
Larger-than-expected accretion disk sizes have also been independently inferred from microlensing observations of gravitationally lensed quasars, where the disk half-light radius (i.e., the characteristic size) is inferred from the differential microlensing variability exhibited by the multiple lensed images~\citep[e.g.,][]{Pooley2007, Dai2010, morgan2010, Jimenez-Vicente2014, Cornachione2020}.
It has been suggested that the measured lags may be contaminated by reprocessed UV/optical photons originating in the broad emission-line region (BLR), which lies farther out in the accretion flow~\citep[e.g.,][]{korista2001, mchardy2018, cackett2018, lawther2018}.
This contamination could cause the observed interband continuum lags to appear longer than those expected from pure disk reprocessing.

In contrast, some studies have argued that accretion disk sizes inferred from CRM are consistent with the standard thin-disk model when factors such as temperature fluctuations within the disk or potential selection biases are properly accounted for~\citep[e.g.,][]{dexter2011a, mudd2018, homayouni2019, yu2020}.
Moreover, \citet{Kammoun2021} showed that the standard thin-disk model can reproduce the observed CRM lags when relativistic (special and general) and disk ionization effects are fully included.
The accretion disk size discrepancy poses one of the significant challenges to the standard thin-disk model and to our understanding of AGN accretion physics more broadly.
A large sample of AGNs with robustly measured CRM lags is required to investigate this issue further.
To date, however, CRM lags have been successfully measured for only a few hundred AGNs~\citep[e.g.,][]{Collier1998, Peterson1998, sergeev2005, jiang2017, homayouni2019, mudd2018, yu2020, guo2022}.

The Vera C. Rubin Observatory's Legacy Survey of Space and Time (LSST) will significantly advance AGN research by providing high-quality multiband light curves for millions of AGNs spanning a wide range of physical properties and redshifts~\citep{ivezic2019}.
LSST's Wide Fast Deep (WFD) survey will monitor 18,000 $\rm deg^2$ of sky with an average cadence of three days in six photometric bands ($ugrizy$), accumulating $\sim\,$800 visits over its 10-year operation.
In addition, LSST will observe five Deep Drilling Fields (DDFs), each covering $\sim$10$\rm\,deg^2$ and receiving at least 10 times more visits than the WFD survey.
Thousands of AGNs in each DDF will be monitored at sub-day cadence (combining all bands) throughout the decade-long operation~\citep{Brandt2018, Kovacevic2022}.
The single-visit 5$\sigma$ limiting magnitudes of LSST reach 23.9, 25.0, 24.7, 24.0, 23.3, and 22.1 in the $u$, $g$, $r$, $i$, $z$, and $y$ bands, respectively~\citep{ivezic2019}.

The high-quality light curves provided by LSST will enable robust characterization of AGN UV/optical variability and interband continuum lags. 
Interband lags can be measured for about one million LSST AGNs~\citep[][Li et al. 2025 submitted]{yu2020a}, allowing detailed investigations of the accretion disk size problem as a function of AGN properties such as $L_{\rm Edd}$ and $M_{\rm BH}$, and offering new insights into the physical origin of continuum lags.
Meanwhile, stochastic variability parameters---such as characteristic timescales and amplitudes---extracted from LSST light curves will enable reliable AGN classification~\citep[e.g.,][]{kozlowski2010, butler2011, macleod2011, peters2015, DeCicco2021, Sanchez-Saez2021, savic2023, Sanchez-Saez2023}. For example, variability-based methods are more effective than SED-based methods at identifying low-luminosity and low-mass AGNs (e.g., in dwarf galaxies), where host-galaxy starlight contamination is significant~\citep[e.g.,][]{baldassare2017, ward2022, bernal2025}. 
A complete census of low-luminosity is essential for understanding AGN evolution across cosmic time~\citep{richards2006a}, while a comprehensive inventory of low-mass AGNs is critical for constraining the formation pathways of SMBHs~\citep[see][and references therein]{greene2020}.

However, existing AGN light curve analysis tools are not well-prepared to fully exploit the scale and richness of LSST observations.
Current tools/methods used to characterize AGN UV/optical variability typically operate on single-band light curves and lack the ability to incorporate multiband information~\citep[e.g.,][]{Vaughan2003, kelly2009, kelly2014, kozlowski2016, yu2022}.
This limitation prevents the effective utilization of LSST's multiband coverage for improved characterization of AGN variability.
Tools/methods commonly used to measure interband lags generally fall into two categories.
The first includes techniques that estimate lags by computing the cross-correlation function (CCF) between light curves; the CCF reaches its maximum when one light curve is shifted relative to the other by an amount corresponding to the intrinsic lag~\citep[e.g.,][]{gaskell1987, edelson1988, Alexander1997}.\footnote{\edit{When applied to a single-band light curve and its shifted copies, the CCF method yields its autocorrelation function (ACF), which can be used to characterize AGN variability~\citep[e.g.,][]{Raiteri2021a, Raiteri2021b}.}}
These methods are typically computationally efficient, but are limited to lag estimation and do not provide a full characterization of the underlying variability process.
The second consists of methods that model each observed light curve as the convolution of an unknown driving continuum with a time-lag distribution, represented by a transfer function~\citep[e.g.,][]{zu2011, li2016, starkey2016, donnan2021}.
This is expressed as,
\begin{equation}\label{eqn:transfer_function}
    y(t) = \int d\tau\, \Psi(\tau)\,y_{\rm drive}(t - \tau),
\end{equation}
where $y_{\rm drive}(t - \tau)$ denotes the flux of the underlying driving continuum at $t - \tau$, and $\Psi(\tau)$ represents the corresponding transfer function.
\jl~is a representative tool in this category~\citep{zu2013}.
It models the driving continuum as a DRW process and adopts a top-hat profile for the transfer function.
While such methods provide a more comprehensive treatment by jointly modeling variability and interband lags, they are often computationally demanding.

Many recent investigations have turned to machine learning (ML) techniques to develop more scalable methods for analyzing AGN light curves~\citep[e.g.,][]{fagin2024, li2024}.
In these approaches, deep neural networks are trained on simulated light curves and are tasked with recovering the underlying variability characteristics and/or the physical parameters used to generate the simulations.
However, the reliability of such methods hinges on the assumption that the models used to generate the training data accurately represent both the variability physics and the observational conditions (e.g., cadence, photometric uncertainty)---an assumption rarely satisfied in practice.
For example, \cite{li2024} show that the AGN physical parameters (e.g., \mbh) predicted by their simulation-based inference model~\citep{cranmer2020} can be biased by up to 0.5 dex when the test light curves are five times noisier than those used for training.

In this paper, we present \pkg---a scalable and flexible AGN light curve modeling tool that leverages LSST's multiband observations to simultaneously infer the properties of the underlying stochastic variability and the lags between photometric bands~\citep{eztaox}.
The name of \pkg~combines three elements: ``Ez", a homophone of easy; ``Tao" referencing the Chinese philosophical and religious tradition of Taoism, commonly translated as ``the Way"; and ``X" denoting that the code is implemented in JAX---a high-performance numerical computing library developed by Google. 
\pkg~achieves a speed increase of $\sim10^2-10^4\times$ on CPUs compared to existing tools with similar capabilities~(e.g., \jl),  making it feasible to perform both stochastic variability characterization and interband continuum lag measurements for all AGNs discovered by LSST. 
Furthermore, \pkg's multiband capability enables robust constraints on stochastic variability timescales (e.g., $\tau_{\rm DRW}$) down to one hour with the LSST WFD cadence, facilitating the identification of low-mass AGNs---such as those residing in dwarf galaxies---which are expected to exhibit characteristic $\tau_{\rm DRW}$ ranging from hours to days~\citep{burke2021, Wang2023a}.

The rest of the paper is organized as follows.
Section~\ref{sec:model} describes the methods and algorithms implemented in \pkg~for characterizing AGN stochastic variability and interband continuum lags.
Section~\ref{sec:results} presents the performance of \pkg~in terms of computational scalability and its robustness in recovering stochastic variability parameters and interband lags. 
In Section~\ref{sec:discuss}, we discuss the scientific opportunities enabled by \pkg, limitations of the current implementation, and potential directions for future development.
Finally, we summarize our findings and conclude in Section~\ref{sec:conclusion}.

\section{Light Curve Modeling Method \& Implementation}\label{sec:model}

\pkg~models AGN UV/optical stochastic variability as a Gaussian process (GP), a method commonly adopted in AGN variability analysis~\citep[e.g.,][]{Press1992, kelly2009, kozlowski2010, Wilkins2019, Griffiths2021, yu2022b, stone2022, Zhang2023, McLaughlin2024}. 
\edit{Since the flux light curves of accreting compact objects typically follow a lognormal distribution~\citep[e.g.,][]{uttley2001, gaskell2004, Giebels2009, macleod2012, decicco2022}, we emphasize that GP modeling is best applied to magnitude light curves, which follow a Gaussian distribution. Moreover, \cite{Gurpide2025} demonstrated that even when the light curves follow a lognormal distribution, the simulated GP parameters can be recovered without significant bias.}

Given a single band AGN light curve, a GP treats the intrinsic brightness of the AGN at each point in time as a Gaussian random variable and captures the temporal correlations between observations through a kernel function $k(t_i, t_j)$, which defines the covariance between the Gaussian variables at any two timestamps $t_i$ and $t_j$~\citep{Rasmussen2006, aigrain2023}.

The likelihood function of a GP given a set of observations $\bf{y}$ is,
\begin{eqnarray}\label{eqn:gp_ll}
\ln \mathcal{L} && = -\frac{1}{2} (\mathbf{y} - \boldsymbol{\mu})^\top \mathbf{K}^{-1} (\mathbf{y} - \boldsymbol{\mu})\\\nonumber
                &&\quad\, - \frac{1}{2} \ln \det(\mathbf{K}) - \frac{N}{2} \ln(2\pi)
\end{eqnarray}
where $\boldsymbol{\mu}$ is the mean vector with its entries equal to the mean of $\bf{y}$, and $N$ is the total number of observations in $\bf{y}$. $\mathbf{K}$ is the GP covariance matrix, with each entry given by
\begin{equation}\label{egn:gen_kernel}
    K_{ij} = k(|t_j - t_i|) + \sigma_{i}^2\,\delta_{ij},
\end{equation}
where $k$ is the aforementioned kernel function, and $t_{i}$ and $t_{j}$ are the timestamps of the $i^{\mathrm{th}}$ and the $j^{\mathrm{th}}$ observation, respectively. The parameter $\sigma_{i}$ denotes the measurement uncertainty of the $i^{\mathrm{th}}$ observation, and $\delta_{ij}$ is the Kronecker delta.
The widely-adopted DRW model can expressed as a GP with the following kernel function:
\begin{equation}\label{eqn:drw_kernel}
    k_{\rm DRW}(|\Delta t|) = \sigma_{\rm DRW}^2\,e^{-|\Delta t|/\tau_{\rm DRW}},
\end{equation}
where $\Delta t$ gives the time difference between any two observations in a light curve.

\pkg~performs multiband modeling and interband lag measurements by treating mean-subtracted light curves from different photometric bands as scaled and time-shifted realizations of a shared underlying GP.
This approach is mathematically equivalent to adopting a Dirac delta function as the transfer function in Equation~\ref{eqn:transfer_function}, with the delta function centered at a constant temporal offset. 
Under this framework, light curves from different photometric bands and different photometric surveys can be combined and modeled using a one-dimensional GP. 
This approach can in principle accommodate any number of photometric bands, although the number of free parameters grows with the number of distinct bands.

The corresponding kernel function for any two observations in such a merged light curve is, 
\begin{equation}\label{eqn:mb_kernel}
    k(|t_j - t_i|) = S_{1}S_{2}\,k_{\rm latent}(|t_j - t_i - \tau_{\rm lag}|),
\end{equation}
where $k_{\rm latent}$ denotes the kernel of the underlying latent GP.
$S_{1}$ and $S_{2}$ are the variability-amplitude scaling factors for the corresponding photometric bands, and $\tau_{\rm lag}$ represents the relative time lag between them. 
The variability amplitude in each band is obtained by multiplying the latent GP amplitude by the corresponding scaling factor.
This class of kernel functions, as expressed in Equation~\ref{eqn:mb_kernel}, has seen broad application in astronomical time series analysis~\citep[e.g.,][]{gordon2020, Villar2021}.
To evaluate the likelihood of the multiband model for a given light curve, we shift the light curve in different bands by their trial $\tau_{\rm lag}$ relative to the reference band, and compute Equation~\ref{eqn:gp_ll} using a kernel that assumes no intrinsic lag between bands (i.e., setting $\tau_{\rm lag} = 0$ in Equation~\ref{eqn:mb_kernel}). 
In this way, the GP parameters and the interband lags are fitted jointly.

\pkg~is built on top of \tp\footnote{\url{https://github.com/dfm/tinygp}}---a GP modeling framework developed using \texttt{JAX}.\footnote{\url{https://github.com/jax-ml/jax}}
\texttt{JAX} is a high-performance numerical computing library that combines automatic differentiation, just-in-time (JIT) compilation, and hardware acceleration to empower scalable machine learning and scientific computing.
\tp~leverages these features of \texttt{JAX} and re-implements the novel \texttt{celerite} algorithm introduced by \cite{foreman-mackey2017} in \texttt{JAX}.
When a kernel function can be expressed as a mixture of exponentials, the \texttt{celerite} algorithm can evaluate the GP likelihood function with $\mathcal{O}(N)$ computational scaling, a significant efficiency enhancement over the standard $\mathcal{O}(N^3)$ scaling associated with direct likelihood evaluation~\citep{foreman-mackey2017}.
\pkg~takes advantage of the \texttt{JAX}-based implementation of the \texttt{celerite} algorithm to provide a scalable solution for multiband light curve modeling of AGNs.

PSD shapes exhibited by the light curves of accreting black holes can be effectively modeled with \texttt{celerite} GP kernels.
For example, the continuous-time autoregressive moving-average (CARMA) process---commonly used as a more flexible alternative to the DRW process---has an equivalent \texttt{celerite} implementation~\citep{rouxalet2002, kelly2014, kasliwal2017, moreno2019, yu2022b}.
A CARMA($p$, $q$) process combines an autoregressive process of order $p$ with a moving-average process of order $q$, and its PSD can be expressed as a weighted sum of modified Lorentzians.
The DRW process is a special case of a CARMA process with $p = 1$ and $q = 0$.
For a detailed description of CARMA modeling, we refer the reader to \cite{kelly2014}.
More recently, \cite{Lefkir2025} introduced an algorithm for approximating arbitrary broken PL PSDs using a sum of multiple \texttt{celerite} kernels, thereby enhancing the interpretability and flexibility of GP-based modeling for AGN light curves.
Moreover, \texttt{celerite}-based kernels can be freely combined through addition and multiplication to model more complex covariance structures, while retaining the $\mathcal{O}(N)$ computational scaling.
All such kernels can be used for simultaneous multiband variability characterization and interband lag measurements.

\begin{deluxetable*}{lccccc}\label{tab:sim_params}
    \tablecaption{Input Parameters for Simulated LSST WFD AGN Light Curves}
    \tablehead{\colhead{Light Curve Set} & \colhead{$g$-band \taudrw} & \colhead{$g$-band \sigmadrw} & \colhead{Interband Lag} & \colhead{Band} & \colhead{Duration}\\
                &   [day] & [mag] & [day] & & [year]}
    \startdata
    One     &           100         &           0.112           & Uniform(0.01, 10) & $g$, $r$ & 3, 10\\
    Two     & LogUniform(0.01, 500) &           0.112           & 0.01$\times$\taudrw      & $u$, $g$, $r$  & 3\\
    Three   &           100         &  LogUniform(0.02, 0.2)    & 1                 & $u$, $g$, $r$  & 3 \\
    \enddata
\end{deluxetable*}
\vspace{-\parskip}

\vspace{-.5cm}
\section{\pkg~Performance}\label{sec:results}
In this section, we evaluate the robustness and scalability of \pkg~and compare it to the widely adopted tool \jl.
Since \jl~supports only the DRW kernel for modeling the driving continuum variability during lag measurement, all performance experiments conducted in this work assume a DRW model for the underlying AGN continuum stochastic variability.
Section~\ref{subsec:robustness} demonstrates \pkg's robustness in recovering input interband lags and DRW parameters from simulated LSST WFD light curves.
\edit{In Section~\ref{subsec:ngc5548} and Section~\ref{subsec:ztf}, we apply \pkg~to continuum reverberation mapping light curves of real AGNs and compare the resulting lag measurement with values reported in the literature.}
Finally, Section~\ref{subsec:scalability} presents computational benchmarking results for \pkg.

\subsection{Robustness: A Test Using Simulated LSST WFD Light Curves}\label{subsec:robustness}

In this section, we simulate realistic LSST WFD light curves, and assess how well \pkg~recovers both the simulated interband lags and stochastic variability parameters.
Section~\ref{subsubsec:lc_sim} describes the light curve simulation process and the sets of input parameters (see Table~\ref{tab:sim_params}).
In Section~\ref{subsubsec:lag_recover}, we show that \pkg~is capable of retrieving interband lags with accuracy comparable to that of \jl.
Section~\ref{subsubsec:drw_tau_recover} and Section~\ref{subsubsec:drw_amp_recover} demonstrates the robustness of \pkg~in recovering input stochastic variability parameters (i.e. \taudrw~and~\sigmadrw).
We emphasize that the goal of the experiments in this section is to demonstrate the overall capability of \pkg, rather than to perform a full parameter-space study or to investigate its systematics. 
Nonetheless, the code for generating and fitting simulated LSST light curves will be made available upon request, allowing interested reader to further explore the robustness of \pkg~under different combinations of input parameters. 

\subsubsection{Light Curve Simulation}\label{subsubsec:lc_sim}

We adopt the DRW model to simulate AGN light curves in individual LSST bands.
We start by simulating DRW light curves in each LSST band with a 30-second cadence; this 30-second cadence is chosen to match the 30-second exposure time of each LSST single visit~\citep{ivezic2019}.
The input DRW parameters across different bands are correlated according to the wavelength-dependent power-law relations presented in \cite{macleod2010}. More specifically,
\begin{eqnarray}\label{eqn:drw_wlen_scale}
    \tau_{\rm DRW} \propto \lambda^{0.17},\quad \sigma_{\rm DRW} \propto \lambda^{-0.479},
\end{eqnarray}
where $\lambda$ is the effective wavelength of the LSST filters~\citep{ivezic2019}. 
We then simulate continuum lags by offsetting light curves in redder band in time with respect to light curves in bluer bands.
Finally, we down-sample the simulated 30-second cadence light curve to the observing cadence of the WFD survey. 

The observational noise is simulated by adding random values drawn from Gaussian distributions with mean zero and variance $\sigma^2_{\mathrm{LSST}}$. The parameter $\sigma_{\mathrm{LSST}}$ is the expected photometric error of LSST, and it can be approximated using
\begin{equation}
    \sigma^2_{\mathrm{LSST}} = \sigma^2_{\mathrm{sys}} + \sigma^2_{\mathrm{rand}},
\end{equation}
where $\sigma_{\mathrm{sys}}$ is the systematic photometric error and $\sigma_{\mathrm{rand}}$ is the random photometric error. LSST aims to achieve a $\sigma_{\mathrm{sys}} < 0.005$ mag~\citep{ivezic2019}; we set $\sigma_{\mathrm{sys}}$ to $0.005$ mag in our simulation. 
The random photometric error ($\sigma_{\mathrm{rand}}$) is expected to scale as
\begin{equation}
    \sigma^2_{\mathrm{rand}} = (0.04 - \gamma)x + \gamma x^2 \, (\mathrm{mag}^2), \quad x = 10^{0.4(m - m_5)},
\end{equation}
where $m_5$ is the 5$\sigma$ limiting magnitude in a given band (for point sources), and $\gamma$ is a band-dependent factor~\citep{ivezic2019}. Following \cite{ivezic2019}, we set $\gamma$ to 0.038 for $u$-band and to 0.039 for other LSST bands.

\begin{figure*}
    \centering
    \includegraphics[width=0.45\linewidth]{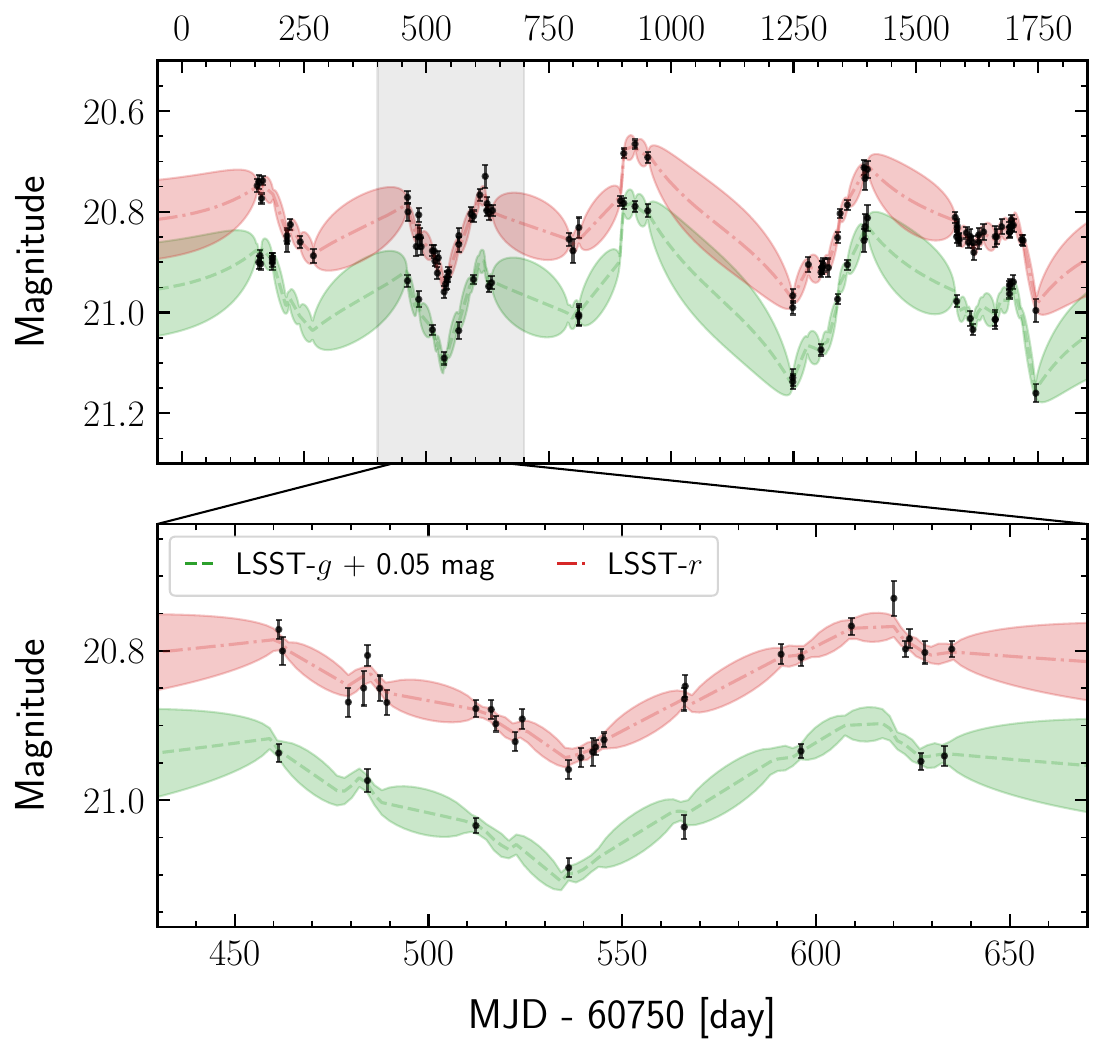}
    \hspace{0.2cm}
    \includegraphics[width=0.52\linewidth]{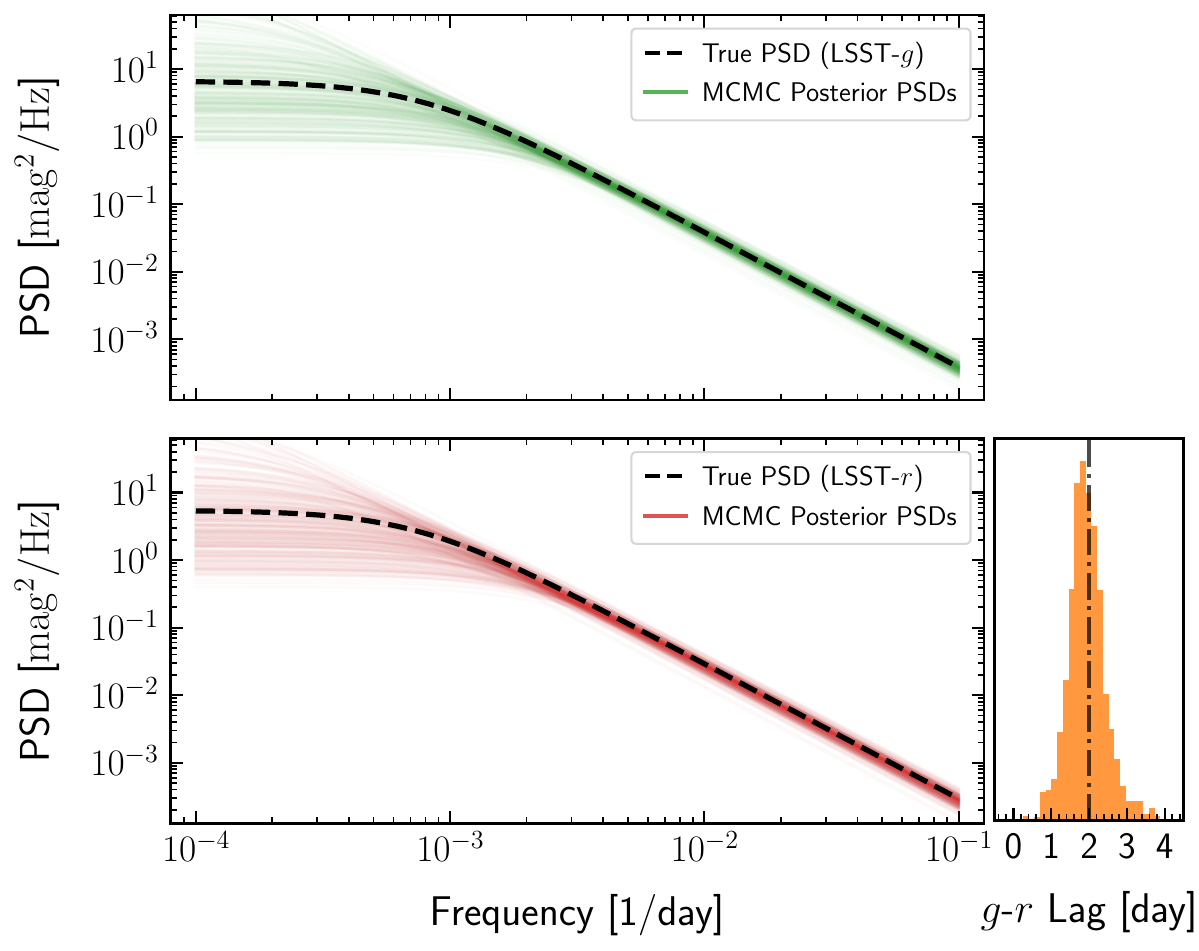}
    \caption{{\it Left}: Example LSST WFD light curves simulated using the DRW model. The $g$- and $r$-band light curves have fiducial mean magnitudes of 21 and 20.9, respectively. 
    {\it Right}: The true (dashed curves) and recovered (colored solid curves) PSDs. The recovered PSDs are generated using random draws of the MCMC posterior distribution. The orange histogram in the bottom right panel shows the posterior distribution of the recovered $g$--$r$ lag, with the dash-dotted vertical line indicating the true simulated lag of 2 days.}
    \label{fig:drw_lc}
\end{figure*}

The observing cadence and the expected $m_5$ at any location on the sky are predicted by LSST operation simulations (\texttt{OpSims}) using MAF~\citep{marshall2017}.
In this work, we adopt the operation simulation version \texttt{baseline\_v4.0\_10yrs}.
All our simulated AGNs also assume a fiducial $g$-band mean magnitude of 21 ($\sim1\%$ photometric accuracy) and mean $u$--$g$ and $g$--$r$ colors of 0.1 mag~\citep{richards2001, temple2021}. 
An example light curve from the first set is shown in Figure~\ref{fig:drw_lc}.

We generate three sets of multiband LSST light curves using different input distributions for the interband lags and DRW parameters. The distributions of input parameters are also summarized in Table~\ref{tab:sim_params}. 
Each set consists of 975 simulated light curves with their cadence and $m_5$ retrieved from 975 unique locations uniformly sampled in the WFD footprint. To construct this sample, we first generate a uniform grid with 5$^\circ$ spacing in both RA and Dec, spanning 0$^\circ$ $<$ RA $<$ 360$^\circ$ and $-75^\circ$ $<$ Dec $<$ $+15^\circ$. We then select grid points that receive between 600 and 1000 visits in all six photometric bands.
The final simulated full 10-year light curves have median numbers of 53, 65, and 171 epochs in the $u$, $g$, and $r$ bands, respectively. When only the first three years of data are considered, these numbers drop to 13, 19, and 50 epochs.

The first set aims to evaluate \pkg's ability to recover interband lags (see Section~\ref{subsubsec:lag_recover}), and the light curves consist of simulated LSST observations in $g$ and $r$ bands. The $g$--$r$ lag for each multiband light curve is uniformly sampled between 0.01 day and 10 days, while the input $g$-band \taudrw~and \sigmadrw~are fixed at fiducial values of 100 days and 0.112 mag, respectively. These choices of \taudrw~and \sigmadrw~are representative of those measured from typical quasar \edit{(i.e., high luminosity AGN)} UV/optical light curves~\citep[e.g.,][]{vandenberk2004, macleod2010}. 

The second set tests \pkg's ability to recover the DRW characteristic timescale \taudrw~(see Section~\ref{subsubsec:drw_tau_recover}) , and the light curves consist of simulated LSST observations in $u$, $g$, and $r$ bands. The input $g$-band \taudrw~is randomly sampled from a log-uniform distribution between 0.01 day and 500 days, while the $g$-band \sigmadrw~is set to a fiducial value of 0.112 mag. The $u$-$g$ and $g$--$r$ lags are set to 1\% of the $g$-band \taudrw. 

The third set evaluates \pkg's ability to recover the DRW asymptotic variability amplitude \sigmadrw~(see Section~\ref{subsubsec:drw_amp_recover}), and the light curves consist of simulated LSST observations in $u$, $g$, and $r$ bands. The input $g$-band \sigmadrw~is randomly sampled from a log-uniform distribution between 0.02 mag and 0.2 mag, while the $g$-band \taudrw~is set to a fiducial value of 100 days. The $u$-$g$ and $g$--$r$ lags are set to 1 day. 


\subsubsection{Recovery of Interband Lag}\label{subsubsec:lag_recover}

\begin{figure*}
    \centering
    \includegraphics[width=0.455\linewidth]{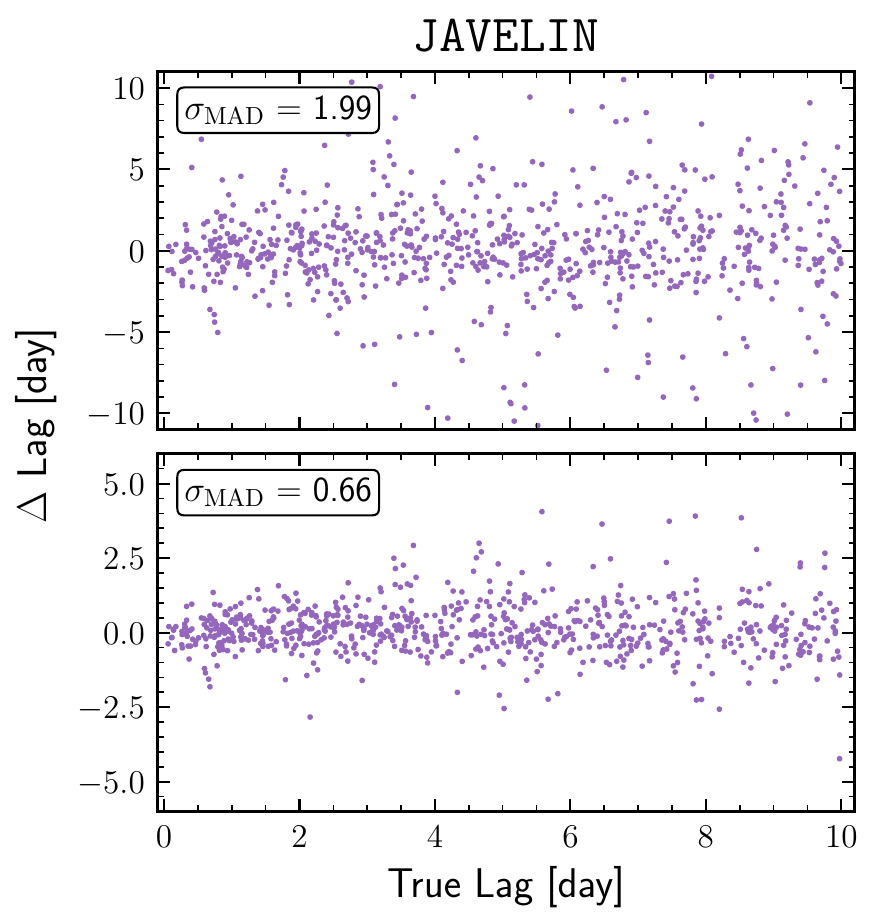}
    \hspace{0.1cm}
    \includegraphics[width=0.477\linewidth]{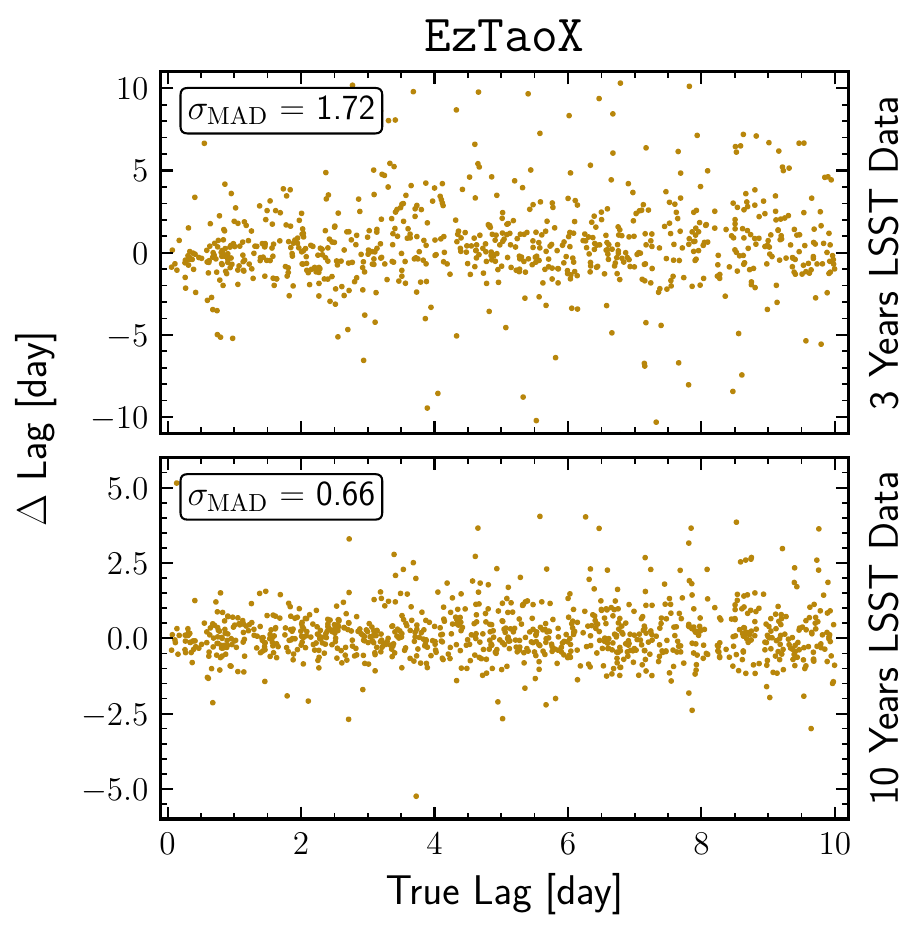}
    \caption{Comparison of $g$--$r$ lags recovered using \jl~(left) and \pkg~(right) from simulated LSST WFD light curves. Both \jl~and \pkg~yield comparable results in terms of lag measurement accuracy. {\it Top:} The difference between the recovered and true lags ($\Delta\:\rm lag$) as a function of the true lag, based on fits obtained using only the first 3 years of WFD observations. 
    The dispersion of $\Delta\,\rm lag$, computed using the median absolute deviation (MAD), is indicated in the top-left corner of each panel.
    {\it Bottom:} Same as the top panels, but using the full 10 years of WFD observations for inference.}
    \label{fig:lag_recover}
\end{figure*}

{\setlength{\parskip}{0pt}
\setlength{\tabcolsep}{14pt}
\begin{deluxetable}{cc}\label{tab:mcmc_prior}
    \tablecaption{MCMC Prior for \pkg~Parameters }
    \tablehead{\colhead{Parameter} & \colhead{Prior Distribution}}
    \startdata
    \taudrw~[day] & LogUniform($10^{-4}$, $10^4$)\\
    \sigmadrw~[mag] & LogUniform($0.01$, $2$)\\
    $S_{\rm band}$ & LogUniform(${0.1}$, ${10}$)\\
    $\tau_{\rm lag}$ [day] & Uniform($-15$, $25$)\\
    $\mu_{\rm band}$ [mag] & Normal($0$, $0.1$)\\
    \enddata
    \tablecomments{The normal distribution follow the notation Normal($\mu$, $\sigma$), where $\mu$ is the mean, and $\sigma$ is the width.}
\end{deluxetable}}

We run \pkg~on the first set of simulated light curves to extract the lag between the $g$- and $r$-band ($\tau_{\rm gr}$), after subtracting the mean from each single-band light curve.
The \pkg~light curve model includes seven parameters: \taudrw, \sigmadrw, $S_{\rm g}$, $S_{\rm r}$, $\mu_{\rm g}$, $\mu_{\rm r}$, and $\tau_{\rm gr}$. 
Here, $S_{\rm g}$ and $S_{\rm r}$ are the GP kernel scaling factors for $g$- and $r$-band observations (see Equation~\ref{eqn:mb_kernel}), while $\mu_{\rm g}$ and $\mu_{\rm r}$ are the mean magnitudes of the $g$- and $r$-band light curves, respectively.
As part of the \pkg~fitting process, we first compute the maximum likelihood estimates (MLEs) of the model parameters, and use the them to initialize the Markov Chain Monte Carlo (MCMC) sampler for posterior sampling. The adopted MCMC priors are listed in Table~\ref{tab:mcmc_prior} and are fixed throughout the analysis unless stated otherwise.
The prior ranges are chosen according to the light curve properties. For example, the lower limit for \taudrw~is set to $\sim$10 times smaller than the minimum time separation between any two observations, while the upper limit is $\sim$10 times larger than the light curve length.
MCMC sampling is performed using the gradient-based No-U-Turn Sampler~\citep[NUTS;][]{Hoffman2011}, with a single chain run for $10^4$ steps following $10^3$ warm-up steps. 
The posterior median of \taudrw~is taken as the point estimate.

We also apply \jl~to the first set of simulated light curves to measure $\tau_{\rm gr}$.
The \jl~light curve model includes five parameters: \taudrw, \sigmadrw, $s_{\rm gr}$, $w_{\rm gr}$, and $\tau_{\rm gr}$. Here, $s_{\rm gr}$ and $w_{\rm gr}$ describe the scale and width, respectively, of the top-hat transfer function for the $g$--$r$ lag.
\jl~employs a two-step fitting procedure: it first generates a distribution of \taudrw~and \sigmadrw~by fitting the light curve of a chosen band via MCMC sampling, and then uses that distribution as a prior for the DRW parameters while simultaneously fitting both the DRW parameters and the interband lags from the light curves in all bands.
In the first step, the light curve in the bluest band is typically used, as \jl~assumes that redder-band light curves are convolved versions of it; here we use the $g$-band light curve.
For MCMC sampling, \jl~employs the affine-invariant ensemble sampler~\citep{goodman2010} as implemented in \texttt{emcee}~\citep{foreman-mackey2013}.
In the first step we follow the default configuration, while in the second step we run the sampler with 100 walkers, 500 burn-in steps, and 3000 sampling steps. 
We also set the prior bound for \taudrw~to [$10^{-4}$, $10^4$] days and for $\tau_{\rm gr}$ to [$-15$, $25$] days. The posterior median of $\tau_{\rm gr}$ is reported as the final point estimate.

We clean the sample of best-fit \pkg~and \jl~lags by removing those associated with non-convergent MCMC chains.
Specifically, we discard fits in which the MCMC chain for $\tau_{\rm gr}$ yields fewer than 400 effective samples ($n_{\rm eff}$), following the recommendation of \cite{Vehtari2021} for selecting converged chains. 
Approximately 20\% of the 3-year and 7\% of the 10-year \pkg~fits are removed by this criterion, while $\sim$20\% of the \jl~fits are removed from both the 3-year and 10-year data.
Examined as a function of the input $\tau_{\rm gr}$, the fraction of discarded \jl~fits increases with increasing true $\tau_{\rm gr}$; in contrast, the fraction of discarded \pkg~fits remains relatively constant across  the range of input $\tau_{\rm gr}$.
We discuss this effect in more detail in Section~\ref{subsec:discuss_transfer}.

\begin{figure*}
    \vspace{0.2cm}
    \centering
    \includegraphics[width=0.635\linewidth]{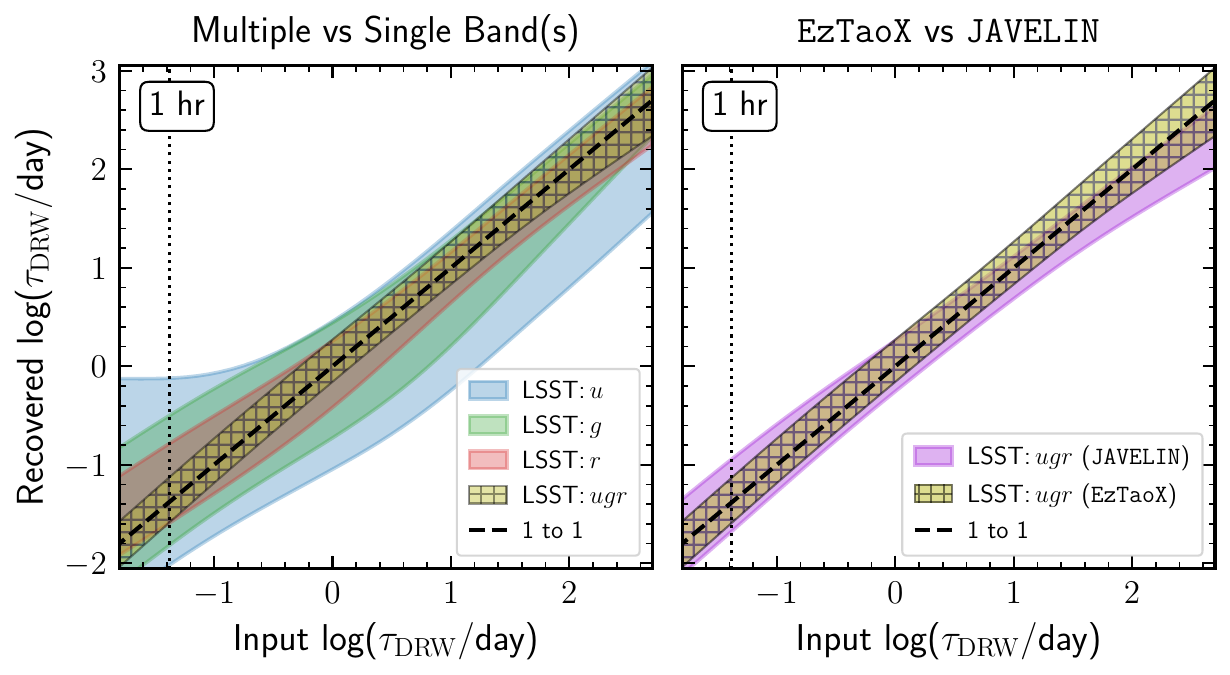}
    \includegraphics[width=0.35\linewidth]{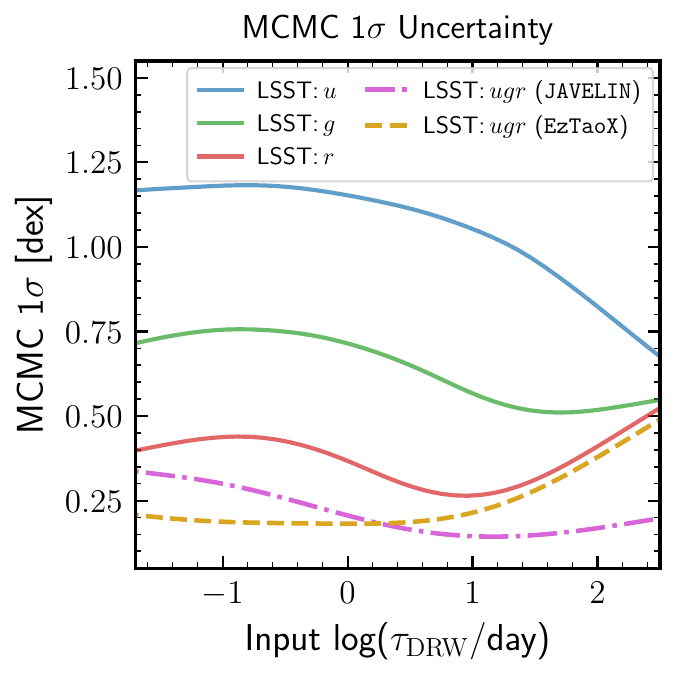}
    \caption{Comparison of recovered and input \taudrw~based on 3 years of simulated LSST WFD observations. \pkg's multiband fitting capability yields more robust and accurate estimates of \taudrw.
    {\it Left}: Comparison of \taudrw~recovered from single-band vs. multiband observations. 
    The blue, green, and red shaded regions represent the central 68\% intervals of the recovered \taudrw~when using only $u$-, $g$-, or $r$-band observations, respectively. 
    The yellow hatched region indicates the corresponding central 68\% interval when observations from all three bands ($u$, $g$, and $r$) are used jointly in the fitting.     
    Using data from all three bands, \pkg~can robustly recover \taudrw~down to 1 hr (marked by the vertical dotted black line), and the associated uncertainty is noticeably reduced in comparison to single-band fits.
    {\it Middle}: Comparison of multiband inference with \pkg~and \jl. The results show that \taudrw~recovered by \jl~are systematically biased when the true log(\taudrw/day) exceeds $\sim1.5$ (\taudrw$\sim30$ days). The purple region denotes the central 68\% intervals of \taudrw~recovered by \jl.
    {\it Right}: The mean MCMC 1$\sigma$ \edit{uncertainty} (defined as one half of the central 68\% credible interval) for \taudrw~demonstrates the improved precision in \taudrw~recovery achieved through multiband fitting. 
    However, the MCMC 1$\sigma$ \edit{uncertainty} from \jl~is clearly underestimated at log(\taudrw/day) $\gtrsim 1.5$, provided with the systematic biases shown in the middle panel.}
    \label{fig:tau_recover}
\end{figure*}

Based on the cleaned samples, \pkg~and \jl~yield comparable results for lag measurement accuracy.
Figure~\ref{fig:lag_recover} presents the difference between the recovered lag and the true simulated lag ($\Delta\rm lag$) as a function of the true lag. 
The left column shows results from \jl, while the right column displays results from \pkg.
The top row corresponds to fits obtained using only the first 3 years of simulated LSST light curves, and the bottom row shows results based on the full 10-year simulated dataset.
The dispersion in $\Delta\rm lag$ is $\lesssim\:$2 days when using 3 years of data and improves to $\lesssim\:$0.7 day with the full 10-year dataset.
These dispersions are consistent with the mean uncertainties derived from the MCMC posterior distributions.

\subsubsection{Recovery of Stochastic Variability Timescale}\label{subsubsec:drw_tau_recover}
We follow the same strategy described in the previous section to fit the second set of simulated light curves.
We apply \pkg~to multiband light curves containing observations in $u$, $g$, and $r$ bands, as well as to light curves with observations in only one of the three bands.
We use the same set of MCMC configurations as those employed in Section~\ref{subsubsec:lag_recover}.
When fitting the multiband light curves, we observed that MCMC chains take significantly longer to converge when the true interband lag is shorter than a few hours. 
This behavior may be driven by the minimum interband cadence (i.e., $u$--$g$, $g$--$r$, etc.) of $\sim40$ minutes, which hampers reliable lag measurements on hourly timescales and below.
To mitigate this issue, and without loss of generality, we fix the interband lags to zero when the MLE of $\tau_{\rm DRW}$ fall below 10 days---corresponding to a simulated lag of $\sim2.5$~hours.
We exclude fits that either exhibit poorly-converged MCMC chains ($n_{\rm eff} < 400$) or yield a best-fit $\tau_{\rm DRW}$ more than an order of magnitude smaller than the minimum cadence.
About 5\% of \pkg~fits are removed by these two criteria. 


Combining observations from multiple bands leads to noticeable improvements in constraints on the DRW characteristic timescale (\taudrw), owing to the enhanced temporal coverage of multiband light curves.
The left panel of Figure~\ref{fig:tau_recover} displays the recovered \taudrw~as a function of the input value. The blue, green, and red shaded regions mark the central 68\% intervals of \taudrw~recovered using $u$-, $g$-, or $r$-band observations, respectively, while the yellow hatched region indicates the interval when all three bands are fitted jointly.
Including all three bands substantially reduces the scatter in the recovered $\tau_{\rm DRW}$.
The right panel of Figure~\ref{fig:tau_recover} further illustrates this gain, showing the mean 1$\sigma$ uncertainty---defined as half the 68\% posterior interval from MCMC---as a function of input $\tau_{\rm DRW}$.
To test the robustness of these uncertainties, we computed the fraction of fits whose 1$\sigma$ \edit{posterior} intervals include the true input $\tau_{\rm DRW}$; for well-calibrated uncertainties this should be close to 68\%. Across input values, the coverage fraction ranges from 60\% to 80\% for both single- and multiband fits. As an additional check, we fit another set of light curves simulated with $g$-band \sigmadrw~$=0.06$ mag and show in Appendix~\ref{appendix:more_eztao_tests} that \pkg~still yields unbiased estimates of \taudrw.

\begin{figure}
    \centering
    \hspace{-0.15cm}\includegraphics[width=.95\linewidth]{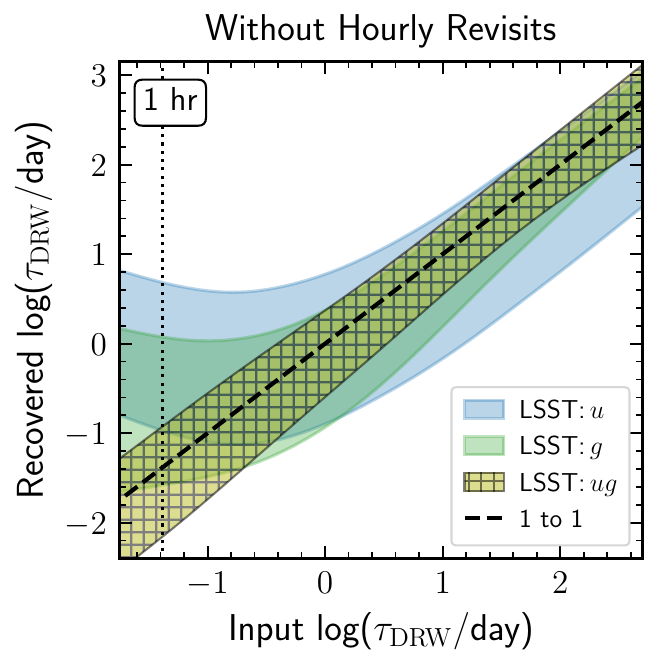}
    \caption{Comparison of recovered and input \taudrw~from simulated LSST WFD light curves without hourly revisits. When the input \taudrw~is less than a few days, estimates recovered using only the $u$- or $g$-band observations show systematic bias. 
    In contrast, combining both $u$- and $g$-band observations yields unbiased \taudrw~estimates down to hourly timescales.}
    \label{fig:tau_recover_hr}
\end{figure}

To better demonstrate the advantages of multiband fitting, we refit the second set of simulated light curves after removing revisits occurring within 1 hr in any given band.
This removal mimics a scenario in which intra-night observations must be stacked to detect faint variable sources against host-galaxy starlight---such as low-luminosity and low-mass AGNs, which are expected to exhibit shorter characteristic variability timescales than typical quasars~\citep[e.g.,][]{burke2021}.
Figure~\ref{fig:tau_recover_hr} displays the recovered \taudrw~as a function of the input value.
When using only $u$- or $g$-band observations, the recovered $\tau_{\rm DRW}$ is systematically biased for input values below a few days. In contrast, combining $u$- and $g$-band data allows $\tau_{\rm DRW}$ to be recovered without bias down to hourly timescales.
Furthermore, MCMC chains fail to converge (i.e., $n_{\rm eff} < 400$) in more than 40\% of the single-band fits, whereas this fraction decreases to only 7\% for the multiband fits, underscoring the improved reliability of multiband fitting.
The multiband capability of \pkg~thus enables robust selection of low-luminosity and low-mass AGNs based on variability, particularly those with intrinsically faint emission and stochastic variability characterized by \taudrw~values of just hours to days.


We also run \jl~on the second set of light curves with observations in all three bands.
To improve convergence, we increase the number of MCMC sampling steps from 3000 to $10^4$.
In the first fitting step, we use the $r$-band light curve to generate the prior distribution of \taudrw~and~\sigmadrw, since the $r$-band provides the densest temporal sampling and lowest photometric uncertainty. 
As shown in Appendix~\ref{appendix:more_jl_tests}, the final DRW parameter distributions recovered by \jl~depend on the photometric band used in the first step, with the $r$-band yielding the most reliable results.
As with \pkg, we find that the MCMC chains exhibit poor convergence when the true interband lag is shorter than a few hours. To mitigate this issue, and without loss of generality, we fix the interband lags to zero when the MLE of $\tau_{\rm DRW}$ falls below 10 days.
Finally, we apply the same sample-cleaning cuts used for \pkg, which remove about 5\% of the \jl~fits.

\jl~produces biased estimates of \taudrw~from our simulated LSST WFD light curves.
The middle panel of Figure~\ref{fig:tau_recover} compares the recovered and true \taudrw. 
The purple region denotes the central 68\% interval of \jl~recoveries, the hatched yellow region shows the corresponding interval for \pkg~recoveries, and the dashed black line marks the one-to-one relation.
For true \taudrw~$\lesssim 3$ days, \jl~slightly overestimates the values, whereas for $\gtrsim 30$ days it underestimates them. 
Moreover, the MCMC-derived uncertainties from \jl~are underestimated for true \taudrw~$\gtrsim 30$ days. The purple dash-dotted curve in the right panel of Figure~\ref{fig:tau_recover} shows the mean 1$\sigma$ interval from the posterior, which is clearly too small in this regime, consistent with the fact that fewer than 50\% of the fits recover the true input within 1$\sigma$.


\subsubsection{Recovery of Stochastic Variability Amplitude}\label{subsubsec:drw_amp_recover}

We fit the third set of light curves with observations in all three bands using both \pkg~and \jl, adopting the same priors and MCMC setup as in Section~\ref{subsubsec:drw_tau_recover}.
We exclude fits that either exhibit poorly-converged MCMC chains ($n_{\rm eff} < 400$) or yield a best-fit value of $\sigma_{\rm DRW} < 0.003$ mag, an order of magnitude below the simulated photometric uncertainty (e.g., $\sigma_{{\rm LSST:}u} \approx 0.03$ mag).
This removes $\sim$5\% of the \pkg~and \jl~fits.

\begin{figure}
    \centering
    \hspace{-0.2cm}\includegraphics[width=.95\linewidth]{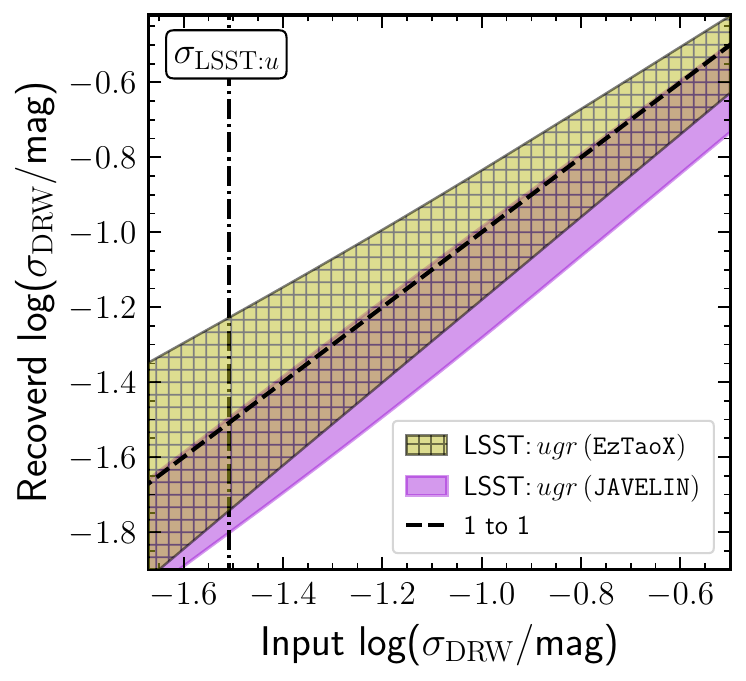}
    \caption{Comparison of recovered and input \sigmadrw~from multiband light curves using $u$-, $g$-, and $r$-band observations with \pkg~and \jl. 
    The hatched yellow region and the purple region represent the central 68\% interval of \pkg- and \jl-inferred \sigmadrw, respectively.
    \pkg~yields unbiased estimates of \sigmadrw~across the full simulated range, from levels comparable to the photometric uncertainty up to $\sim\,$0.2 mag. 
    In contrast, \jl~fails to recover \sigmadrw~when it approaches a few times the $u$-band photometric uncertainty, as marked by the vertical dash-dotted line.
    }
    \label{fig:amp_recover}
\end{figure}

\pkg~is able to recover \sigmadrw~without bias down to levels comparable to the photometric uncertainty of the simulated observations.
Figure~\ref{fig:amp_recover} shows the recovered versus input \sigmadrw.
The hatched yellow region and the purple region denote the central 68\% intervals of \sigmadrw~recovered by \pkg~and \jl, respectively.
Across the full range of simulated \sigmadrw, \pkg~provides unbiased estimates.
In contrast, \jl~systematically underestimates \sigmadrw, with values falling below the one-to-one relation (dashed diagonal). 
We suspect the bias likely arise from the underestimation of \taudrw~seen in the middle panel of Figure~\ref{fig:tau_recover}.
Because \taudrw~and \sigmadrw~are correlated, \jl~yields biased estimates of both parameters when the true \taudrw~exceeds $\sim$30 days in a 3-year light curve.
Finally, to demonstrate the robustness of \pkg~in this regime, we simulate and fit an additional set of light curves with \taudrw~$=200$ days; the results are presented in Appendix~\ref{appendix:more_eztao_tests}.


\begin{figure*}
    \centering
    \hspace{-1cm}\includegraphics[width=0.9\linewidth]{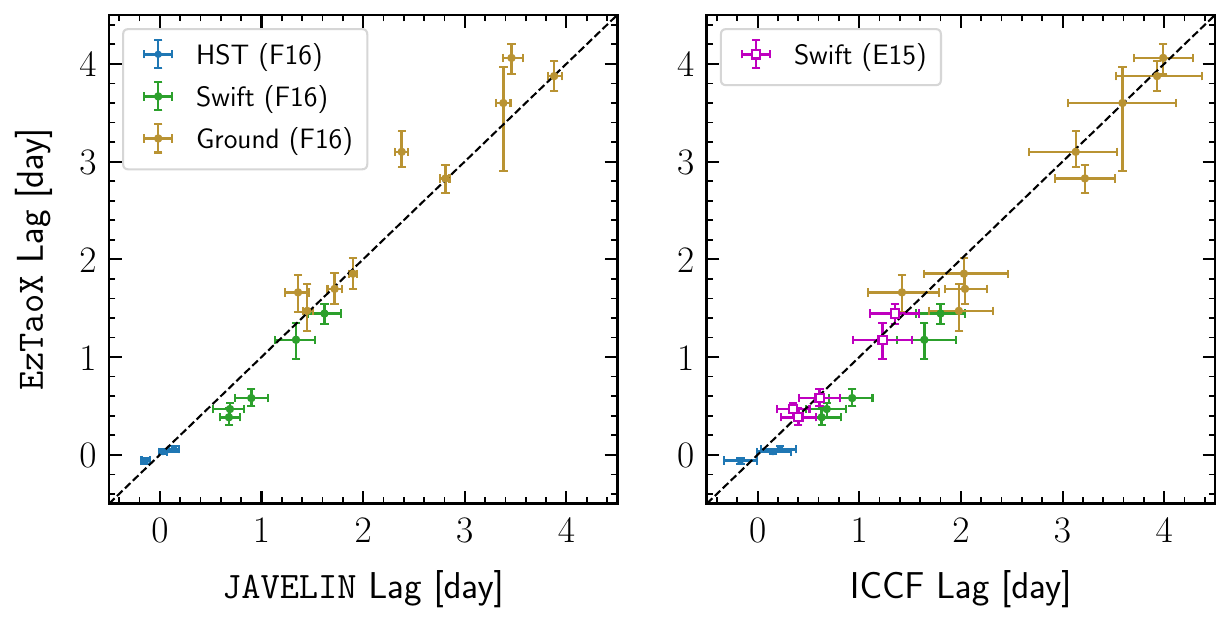}
    \caption{Comparison of interband continuum lags for NGC 5548 measured with \pkg~to those reported in the literature. The light curves of NGC 5548 are obtained by the AGN STORM project. 
    In the legend, F16 refers to the lags reported by \cite{fausnaugh2016}, and E15 refers to the lags reported by \cite{Edelson2015}.
    Interband continuum lags measured with \pkg~are broadly consistent with those obtained using \jl~and the ICCF method, highlighting the effectiveness of \pkg~despite its adoption of a simplified transfer function (i.e., a Dirac delta function).}
    \label{fig:ngc5548}
\end{figure*}

\begin{figure*}
    \centering
    \hspace{-1cm}\includegraphics[width=0.9\linewidth]{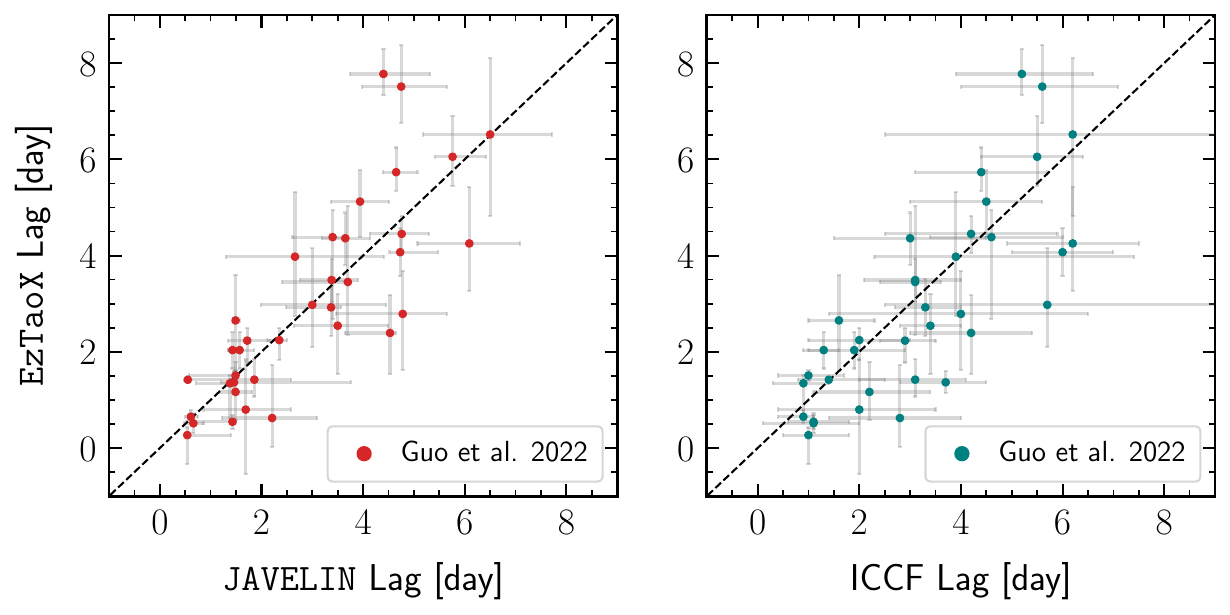}
    \caption{Comparison of interband continuum lags for a sample of type 1 AGNs measured with \pkg~to those reported in the literature~\citep{guo2022}. The light curves are obtained from ZTF. Interband continuum lags measured with \pkg~are broadly consistent with those obtained using \jl~(left) and the ICCF method (right)}
    \label{fig:ztf}
\end{figure*}

\subsection{A Test Using Continuum Reverberation Mapping Light Curves of NGC 5548}\label{subsec:ngc5548}

The AGN Space Telescope and Optical Reverberation Mapping (STORM) project is an intensive multiwavelength monitoring campaign targeting NGC 5548, a well-studied Seyfert galaxy~\citep{DeRosa2015, Edelson2015, fausnaugh2016}.
This campaign combines observations from the \textit{Hubble Space Telescope} (\textit{HST}), the \textit{Neil Gehrels Swift Observatory}~\citep[\textit{Swift} hereafter;][]{Gehrels2004}, and a network of ground-based telescopes, to provide simultaneous daily light curves of NGC 5548 across the X-ray, UV, and optical bands. 
For further details on the observational setup and data, we refer interested readers to \cite{DeRosa2015}, \cite{Edelson2015}, and \cite{fausnaugh2016}.

Applying \pkg~to the UV/optical AGN STORM light curves of NGC 5548 reveals that the lags inferred by \pkg~are broadly consistent with literature values derived using \jl~and ICCF.
We follow the same strategy described in Section~\ref{subsubsec:lag_recover} to measure interband lags relative to the continuum light curve at 1367\AA. In contrast to the approaches taken by \cite{Edelson2015}, and \cite{fausnaugh2016}, we do not de-trend the light curves prior to lag estimation.
We also increased the prior range of $S_{\rm band}$ to Uniform($10^{-5}$, $10$) because AGN STORM light curve are provided in flux, which exhibits large amplitude variation across bands. 

Figure~\ref{fig:ngc5548} compares the lags measured with \pkg~to those derived with \jl~and ICCF. F16 and E15 in the legend refer to the lags reported by \citet{fausnaugh2016} and \citet{Edelson2015}, respectively.
The blue points represent the continuum lags measured at 1158\AA, 1479\AA, and 1746\AA~ using \textit{HST} light curves. The green points correspond to the continuum lags from the \textit{Swift} UVW2, UVM2, UVW1, U, and B bands. 
The yellow points represent continuum lags from Johnson/Cousins \textit{BVRI} bands and Sloan Digital Sky Survey~\citep[SDSS;][]{york2000} $ugriz$ bands \edit{based on ground-based observations.}
The purple squares in the right panel of Figure~\ref{fig:ngc5548} shows the \textit{Swift} lags reported by \cite{Edelson2015}.
We note that the \edit{\it Swift} ICCF lags from \cite{fausnaugh2016} are systematically larger than those reported by \cite{Edelson2015}, likely due to the different de-trending approaches they adopted. 
Overall, interband lags measured using \pkg~are broadly consistent with those obtained using \jl~and ICCF.

\begin{figure}
    \centering
    \hspace{-.2cm}\includegraphics[width=0.99\linewidth]{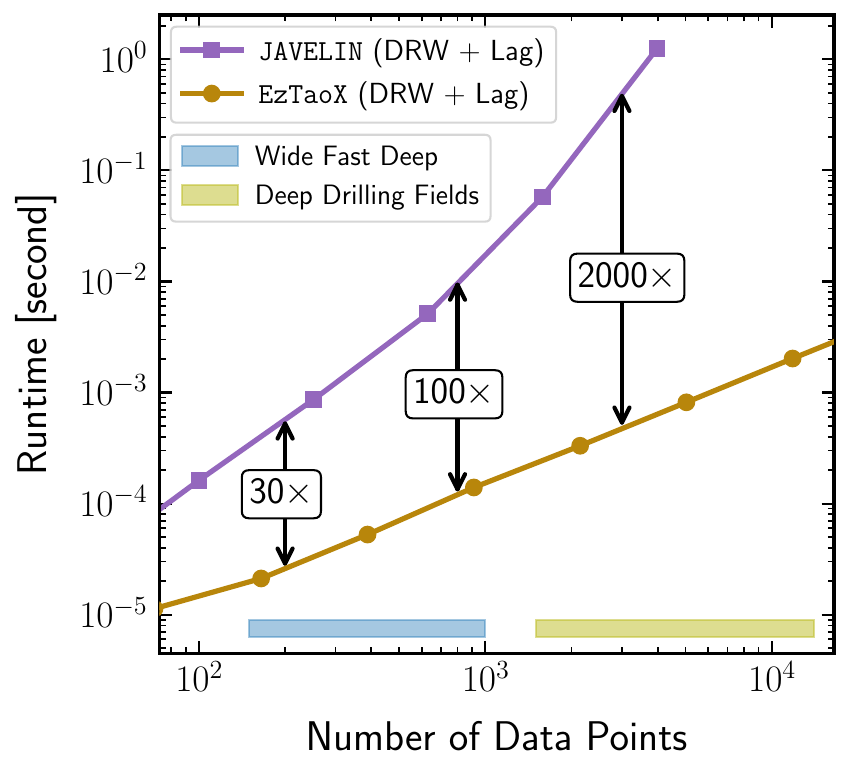}
    \caption{A computational benchmark comparison of \pkg~and \jl~for evaluating the likelihood function of a light curve model that jointly estimates both the DRW parameters and the interband lag.
    The $x$-axis represents the total number of data points in a multiband light curve, while the $y$-axis indicates the corresponding computational cost for a single likelihood evaluation.
    The blue stripe near the bottom marks the parameter space corresponding to WFD light curves, ranging from $\sim\,$200 epochs in the $g$ and $r$ bands to $\sim\,$800 epochs across all six bands.
    The yellow stripe at the bottom marks the parameters space corresponding to DDF light curves. 
    \pkg~is $30$$-$$100\times$ faster than \jl~for WFD light curves and more than 2000$\times$ faster for DDF light curves.
    }
    \label{fig:bm}
\end{figure}

\subsection{A Test Using Zwicky Transient Facility AGN Light Curves}\label{subsec:ztf}

\edit{The Zwicky Transient Facility (ZTF) is an optical time-domain survey that covers the entire sky north of Dec = $-30^\circ$ every $\sim2-3$ nights in the $gri$ bands~\citep{ZTF2019a, ZTF2019b}. The median 5$\sigma$ depths of ZTF reach $g \sim 20.8$, $r \sim 20.6$, and $i \sim 20.0$ mag (AB system) for a 30~s exposure~\citep{Masci2019}.
Using light curves from ZTF Data Release~7 (DR7) , \citet{guo2022} measured continuum lags for a large sample of type~1 AGNs at $z < 0.8$, among which 38 objects (the “core” sample) have high-quality interband lag measurements from both \jl~and ICCF.}

\edit{We apply \pkg~to the $g$- and $r$-band ZTF light curves of the core-sample AGNs from \citet{guo2022}, and find that the lags measured with \pkg~are broadly consistent with those obtained using \jl~and ICCF.
Figure~\ref{fig:ztf} compares the interband lags measured with \pkg~to those from \jl~(left panel) and ICCF (right panel).
We retrieve the ZTF light curves for these sources from the IPAC Infrared Science Archive~\citep{ztf_lightcurve} and truncate them to match the DR7 temporal baseline (i.e., MJD~<~59400). When multiple light curves are available for the same object (each obtained from a different CCD), we select the one with the largest number of epochs.
We follow the same strategy described in Section~\ref{subsubsec:lag_recover} to measure the lags between the $g$ and $r$ bands. We also fit additional jitter terms---one for each band---added to the diagonal elements of the covariance matrix to account for any residual photometric uncertainties not captured by the photometric pipeline~\citep{burke2021}.}


\subsection{Scalability: Computation Time}\label{subsec:scalability}
For LSST light curves, \pkg~is $\sim10-10^3\times$ faster than \jl~per likelihood evaluation.
Figure~\ref{fig:bm} displays the computation time benchmarks for \pkg~and \jl, performed on an Apple Mac Mini with an M2 Pro chip, and the results reflect single-core CPU performance.\footnote{\edit{We also benchmarked \pkg~under multi-core settings (i.e., each core running an independent benchmarking job in parallel), and the resulting performance remained unchanged.}}
For a typical 10-year $g$ and $r$ LSST WFD light curve---with $\sim\,$60 $g$-band observations and $\sim\,$170 $r$-band observations~\citep{ivezic2019}---\pkg~achieves a reduction in computation time of $\sim\,$30$\times$ over \jl. When light curves in all six bands are jointly modeled (assuming a total of 800 observations), the performance gain further increases to $\sim\,$100$\times$.
For a typical 10-year $g$ and $r$ LSST DDF light curve, which contains at least $10$ times more observations in both the $g$ and $r$ bands, \pkg~achieves a speed increase exceeding $2000\times$.
The benchmark results are based on two-band light curves; however, the computational cost depends primarily on the total number of data points in the light curve. 
\edit{We suspect the observed speed increase primarily reflects the algorithmic advantages of \texttt{celerite}; however, without the JIT compilation provided by JAX, \pkg~would be $\sim5-10\times$ slower for WFD light curves.}

An additional increase in speed of up to 10$\times$ can be achieved by using more efficient sampling algorithms/tools.
Upon obtaining a point estimate of the model parameters through optimizing the likelihood function, uncertainties on model parameters can be quantified through posterior sampling, typically performed using MCMC methods.
Most existing AGN light curve modeling tools adopt \texttt{emcee} for MCMC sampling, which is adequate for most inference tasks.
However, \texttt{emcee} can become computationally inefficient when nontrivial or computationally intensive prior distributions are introduced.
\texttt{NumPyro}\footnote{\url{https://github.com/pyro-ppl/numpyro}}---a probabilistic programming library built on \texttt{JAX}---overcomes this limitation by compiling custom prior distribution functions at runtime using JIT compilation, thereby enabling more efficient posterior sampling~\citep{phan2019composable, bingham2019pyro}.
\texttt{NumPyro} provides efficient implementations of various MCMC algorithms, including the affine-invariant ensemble sampler implemented in \texttt{emcee} and NUTS. In our tests, NUTS implemented in \texttt{NumPyro} achieves up to a tenfold speed increase over \texttt{emcee} in generating 5000 effective posterior samples ($n_{\rm eff}$), when applied to the same set of two-band light curves and using identical prior distributions for \pkg~parameters. 


\section{Discussion}\label{sec:discuss}

\subsection{Unbiased Point Estimator of \taudrw}\label{subsec:discuss_drw_tau}

Unbiased recovery of \taudrw~depends on the light-curve cadence, baseline (i.e., the total duration of observations), and adopted priors.
As shown by \citet{kozlowski2017a} and later studies~\citep[e.g.,][]{sanchez-saez2018, burke2021, suberlak2021, Hu2024}, \taudrw~is systematically underestimated when the light-curve baseline is shorter than $\sim$10 times the true \taudrw.
\cite{Hu2024} conducted extensive simulations demonstrating that the assumed prior on \taudrw~also impact its recovery.
In addition, they reported that \taudrw~is systematically overestimated when the true \taudrw~is less than five times the mean cadence, since the \edit{temporal} sampling then probes only the white-noise regime (i.e., a PL exponent of zero) of the DRW PSD.

\begin{figure}
    \centering
    \hspace{-0.1cm}\includegraphics[width=.99\linewidth]{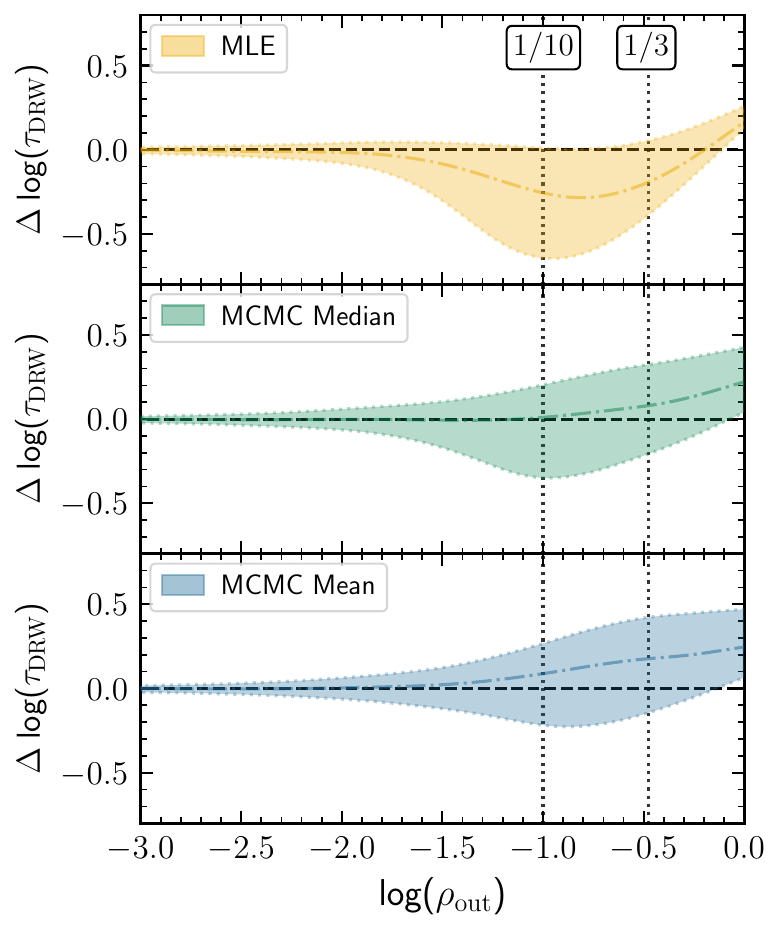}
    \caption{Comparison of three point estimators for \taudrw: MLE, MCMC posterior median, and MCMC posterior mean. 
    The $x$-axis is the recovered \taudrw-to-baseline ratio ($\rho_{\rm out}$), and the $y$-axis is the difference between the recovered and true values of log(\taudrw).
    In each panel, the dash-dotted line marks the median of the $\Delta\,$log(\taudrw) distribution, and the shaded region shows its central 68\% interval.
    The MCMC posterior median is the most reliable, yielding systematically unbiased estimates up to one tenth of the light-curve baseline (vertical dotted line, left) and introducing only a small systematic bias of $\sim0.1$ dex at one third of the baseline (vertical dotted line, right).
    }
    \label{fig:best_fit_select}
\end{figure}

Following previous investigations, we perform simulations to assess how accurately \pkg~recovers \taudrw~as a function of \taudrw-to-baseline ratio.
We simulated 5000 single-band light curves with a uniform 2-day cadence and a baseline of $10^4$ days. Each light curve adopts a fiducial \sigmadrw~of 0.112 mag and a mean photometric uncertainty of 0.01 mag. The input \taudrw~ranges from 10 days to $10^4$ days, corresponding to an input \taudrw-to-baseline ratios ($\rho_{\rm in}$) spanning from 0.001 to 1. 
We fit the simulated single-band light curves using \pkg~to infer \taudrw~and \sigmadrw, following the strategy outlined in Section~\ref{subsubsec:lag_recover}. 
To avoid potential prior-driven biases, we extend the upper limit of the MCMC prior on \taudrw~to $10^5$ days---an order of magnitude longer than the light-curve baseline.

\begin{figure*}
    \centering
    \includegraphics[width=0.44\linewidth]{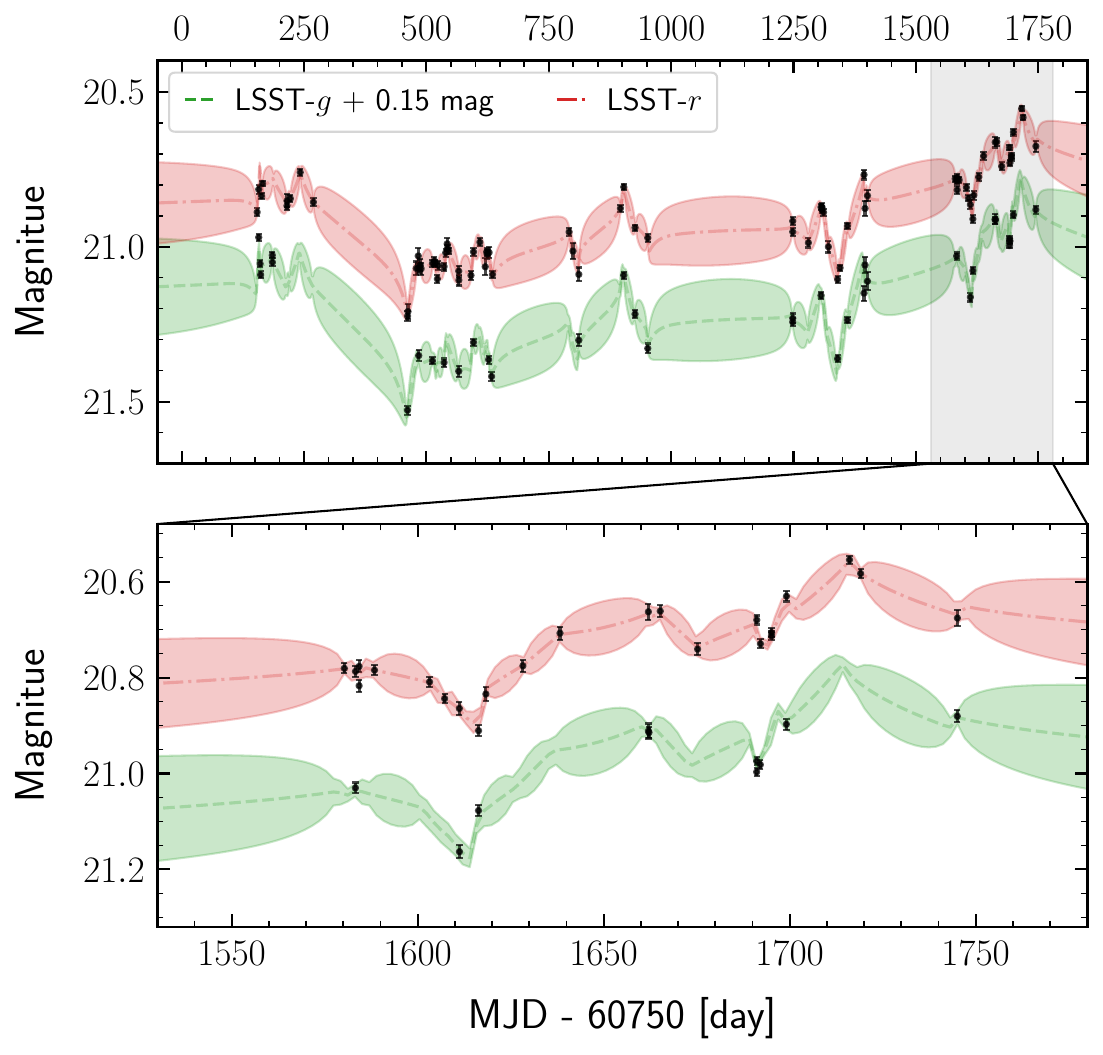}
    \hspace{0.2cm}
    \includegraphics[width=0.525\linewidth]{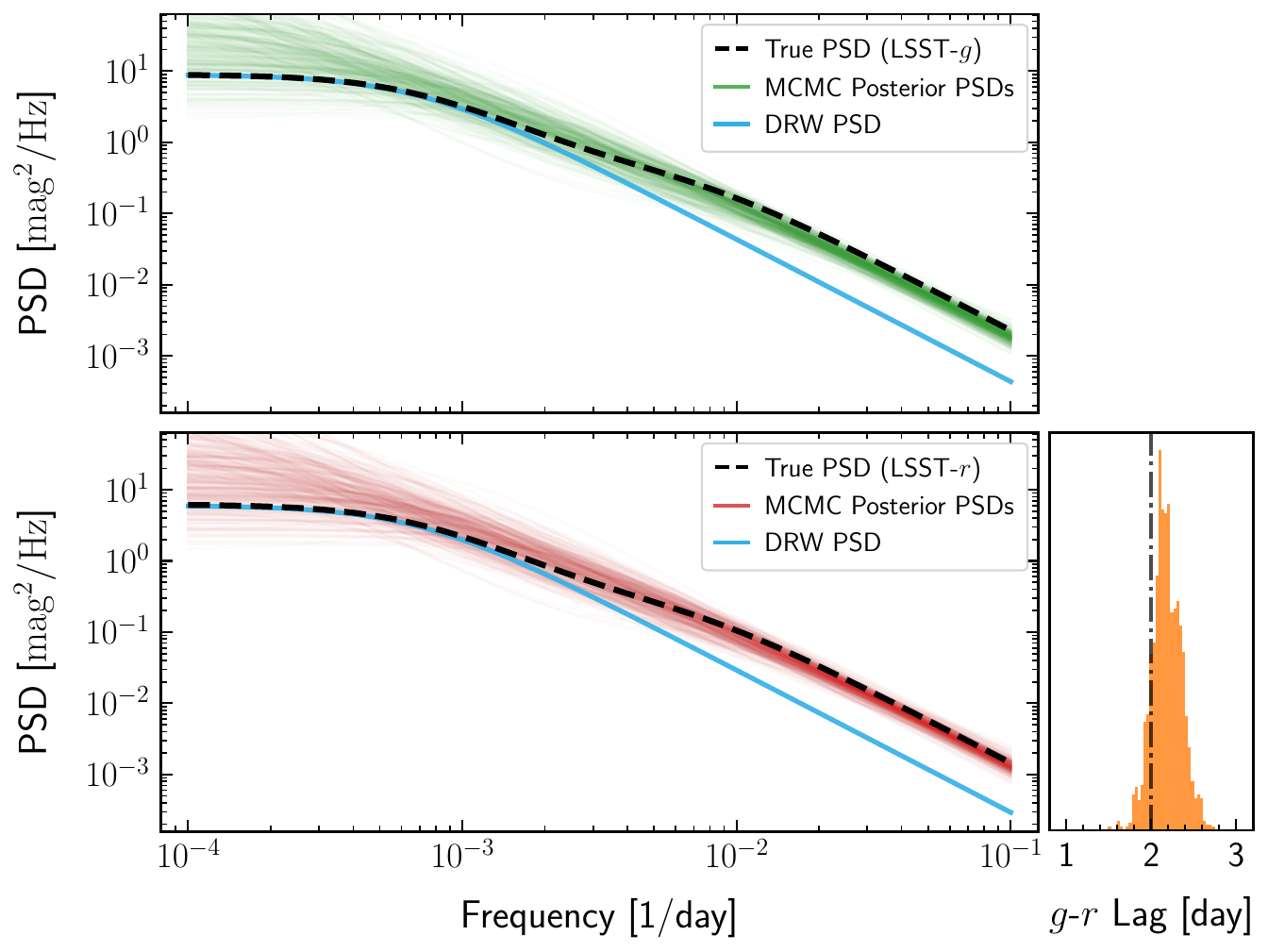}
    \caption{{\it Left}: Example LSST WFD light curves simulated using a CARMA(2,1) model. The $g$- and $r$-band light curves have fiducial mean magnitudes of 21 and 20.9, respectively. 
    {\it Right}: The true (dashed curves) and recovered (colored solid curves) PSDs. 
    The recovered PSDs are generated using random draws of the MCMC posterior distribution. 
    As a comparison, the blue dashed curves display the reference DRW PSDs.
    The orange histogram in the bottom right panel shows the posterior distribution of the recovered $g$--$r$ lag, with the dash-dotted vertical line indicating the true simulated lag of 2 days.}
    \label{fig:dho_demo}
\end{figure*}

Based on our simulations, \pkg~recovers \taudrw~without systematic bias up to one tenth of the light-curve baseline. 
Figure~\ref{fig:best_fit_select} displays the distribution of $\Delta\,$log(\taudrw)---the difference between the recovered and input log(\taudrw)---as a function of the recovered \taudrw-to-baseline ratio ($\rho_{\rm out}$).
Plotting $\rho_{\rm out}$ rather than $\rho_{\rm in}$ is more informative, since the true \taudrw~is unknown when fitting DRW models to real AGN light curves.
In each panel, the dash-dotted line indicates the median of the $\Delta\,$log(\taudrw) distribution, and the shaded region shows its central 68\% interval.
The top, middle, and bottom panels correspond to three common point estimators: MLE, MCMC posterior median, and MCMC posterior mean, respectively.
Among these, the MCMC posterior median is the most robust, remaining systematically unbiased as long as the the recovered \taudrw~is less than one-tenth of the baseline (dash-dotted lines in the middle panel of Figure~\ref{fig:best_fit_select}).
It introduces only a small systematic bias ($\sim0.1$ dex) when the recovered \taudrw~is about one-third of the baseline.

\subsection{Deviation from DRW}\label{subsec:discuss_not_drw}

In this work, we assume that AGN stochastic variability follows a DRW process; however, the modeling framework introduced here is sufficiently flexible to accommodate alternative GP kernels for characterizing the underlying variability.
As an illustration, Figure~\ref{fig:dho_demo} presents $g$- and $r$-band light curves simulated using a CARMA(2,1) process, sampled according to the LSST WFD cadence (left), along with the corresponding PSDs recovered using \pkg~(right). 
The CARMA(2,1) process---also known as the noise-driven damped harmonic oscillator (DHO) model---has been shown to provide superior fits to observed AGN UV/optical light curves compared to the DRW model~\citep[e.g.,][]{kasliwal2017, Yu2025}.
For reference, the DRW PSD is shown in blue in the right panel. 
Similar to Figure~\ref{fig:drw_lc}, the orange histogram in the lower right corner displays the posterior distribution of the $g$--$r$ lag.
The specific CARMA(2,1) process simulated is governed by the following kernel:
\begin{equation}
    k(t_i, t_j) = a_1 e^{-c_1\,|t_j - t_i|} + a_2 e^{-c_2\,|t_j - t_i|},
\end{equation}
which is equivalent to the sum of two DRW kernels.

\subsection{Beyond the Dirac Delta Transfer Function}\label{subsec:discuss_transfer}

\pkg~models interband lags in multiband light curves using a Dirac delta transfer function, a simplified approach compared to methods that adopt more flexible transfer functions with finite width and skewed shapes~\citep[e.g.,][]{zu2011, starkey2016, li2016, donnan2021}.
While more sophisticated transfer functions can provide additional insights into the geometry and physical properties of AGN accretion disks~\citep[e.g.,][]{Cackett2007, starkey2016, fagin2024}, we argue that the cadence of LSST WFD light curves is likely insufficient to resolve the detailed shape of the transfer functions.
Thus, the simplified Dirac delta transfer function adopted by \pkg~is likely adequate for lag estimation from LSST WFD light curves.
Furthermore, previous investigations and the results of Section~\ref{subsec:ngc5548} show that even with high-cadence continuum reverberation mapping data, the use of more flexible transfer functions does not necessarily result in significantly different mean lag estimates compared to simpler methods~\citep[e.g.,][]{Fausnaugh2018, donnan2023}.
An exception perhaps arises when the intrinsic transfer function is notably skewed, in which case simplified methods may underestimate the mean lag~\citep{yu2020a, Chan2020}.

\begin{figure}
    \vspace{0.2cm}
    \centering
    \hspace{-0.2cm}\includegraphics[width=1\linewidth]{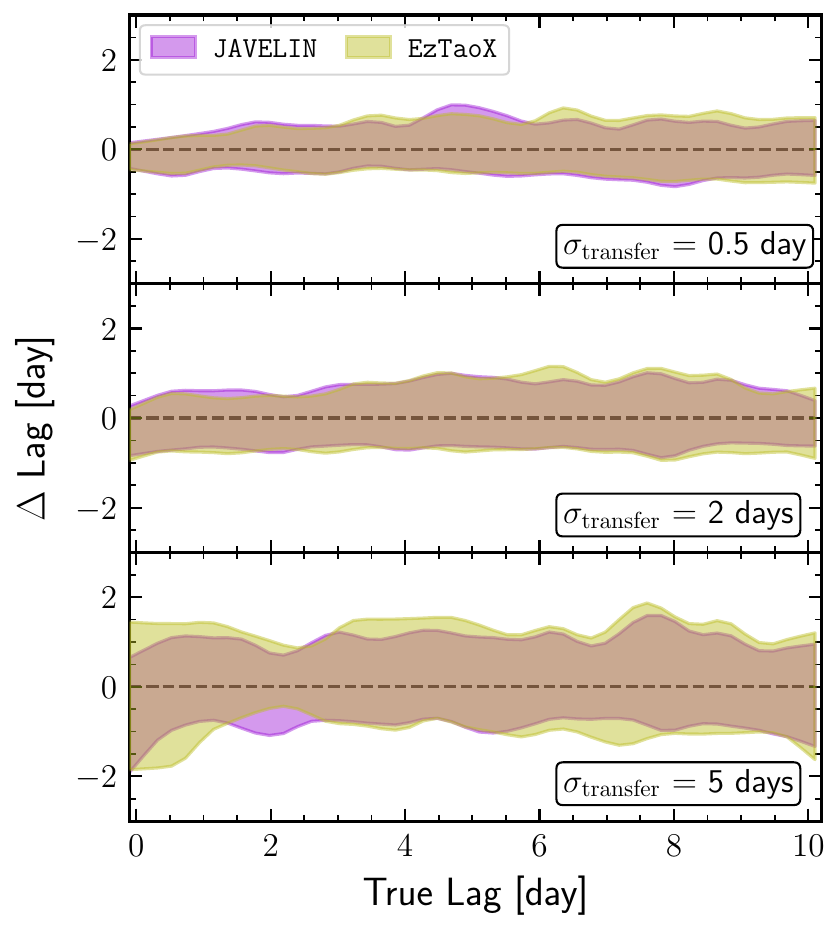}
    \caption{Comparisons of $g$--$r$ lags extracted by \pkg~and \jl~from 10-year LSST WFD light curves simulated with Gaussian transfer functions of width 0.5, 2, and 5 days (top to bottom). 
    The shaded regions shows the central 68\% interval of the $\Delta\,$lag distribution.    
    \pkg~yields lag measurements comparable to those of \jl, even when the light curves are simulated with a transfer function width as wide as 5 days.
    }
    \label{fig:discuss_gauss_lag}
\end{figure}

As an extension to the comparison in Section~\ref{subsubsec:lag_recover}, we test \pkg~and \jl~on new sets of 10-year LSST WFD light curves simulated with a Gaussian transfer function. As before, the $g$-band light curves adopt an input \taudrw~= 100 days and \sigmadrw~= 0.112 mag. The $r$-band light curves are generated by convolving the $g$-band curves with a Gaussian transfer function.
We simulate three sets of light curves with fixed Gaussian widths of 0.5, 2, and 5 days.
Following the same procedure as in Section~\ref{subsubsec:lag_recover}, we extract $g$--$r$ lags from the simulated light curves using both \pkg~and \jl; the results are shown in Figure~\ref{fig:discuss_gauss_lag}. We find that \pkg~yields lag measurements comparable to those of \jl~even when the transfer function width is as large as 5 days. 
Moreover, the $n_{\rm eff}$-based sample cleaning procedure tends to removes a higher fraction of \pkg~fits with true lags much smaller than the transfer function width, while preferentially discarding \jl~fits with true lags much larger than the width. This behavior explains the trend noted in Section~\ref{subsubsec:lag_recover}, where the fraction of discarded \jl~fits increases with true lag when the light curves are simulated with a Dirac delta transfer function.


\subsection{LSST Variability Science Enabled by \pkg}\label{subsec:discuss_science}

\pkg~enables reliable and simultaneous characterization of variability and interband lags for the tens of millions of AGNs to be monitored by LSST.
Existing non-ML methods/tools capable of jointly characterizing stochastic variability and measuring interband lags are not scalable to LSST data volumes~\citep[e.g.,][]{zu2011, starkey2016, donnan2021}.
While ML approaches are scalable, their reliability critically depends on the assumption that the training data accurately capture the true nature of AGN variability---an assumption that is rarely satisfied in practice.
In comparison to traditional non-ML tools, \pkg~achieves a speed increase of $>100\times$ for LSST WFD light curves, and $>10^4\times$ for LSST DDF light curves, without compromising ---and in some cases improving---the accuracy of the results.
Furthermore, \pkg~offers the flexibility to incorporate more expressive GP kernels, allowing for improved modeling of intrinsic continuum variability.
This combination of scalability and flexibility enables robust characterization of variability and interband lags for an unprecedented number of AGNs.
A large, statistically significant sample of AGNs with well-characterized continuum variability and interband lags will facilitate detailed investigations into the fundamental drivers of AGN variability and the physical origins of continuum lags.

\pkg~enables robust variability selection of actively accreting MBHs. 
Accreting MBHs in dwarf galaxies are difficult to identify due to their low luminosity relative to their host galaxy starlight~\citep{greene2020}.
Nevertheless, their characteristic stochastic variability provides a powerful selection tool in the time domain~\citep[e.g.,][]{baldassare2018, ward2022, burke2022, bernal2025}.
By modeling dwarf galaxy image-differencing light curves as a DRW process, one can identify MBH candidates, which are expected to exhibit characteristic \taudrw~values ranging from hours to days~\citep{burke2021, Wang2023a}.
Given the cadence of LSST WFD light curves, the multiband modeling capability of \pkg~enables precise measurement of \taudrw~on timescales from one hour to one day (see Figure~\ref{fig:tau_recover}), outperforming single-band approaches.
When nightly stacked images are required to detect variability from fainter sources, the intra-night cadence in single-band LSST light curves is lost, leading to biased \taudrw~estimates (see Figure~\ref{fig:tau_recover_hr}).
In contrast, through multiband fitting, \pkg~recovers \taudrw~without bias down to hourly timescales for AGNs detected in the WFD survey.

\subsection{Caveats \& Future Work}\label{subsec:discuss_future}

Variable emission from extended re-processors is not explicitly modeled in the current implementation of \pkg.
Recent investigations suggest that reprocessed emission arising at distances much larger than the scale of the accretion disk---for example, in the BLR---can contribute to the measured continuum lags, potentially leading to overestimated accretion disk sizes relative to theoretical predictions~\citep[e.g.,][]{korista2001, mchardy2018, cackett2018, lawther2018, Lira2024, Jaiswal2024, Leighly2024}.
The current multiband light curve model in \pkg~assumes a single lag between any two photometric bands, and therefore lacks the capacity to disentangle multi-component lag distributions, such as reprocessed emission arising from the accretion disk versus that from much larger scales.
Developing methodologies that can decompose observed light curves into distinct lagged components would be highly valuable---not only for probing the detailed geometry of AGN accretion flows, but also for advancing our understanding of the physical drivers behind AGN stochastic variability.

The GP kernels currently supported by \pkg~are all stationary; however, AGN stochastic variability may exhibit non-stationary behavior.
A notable example is the class of changing-look AGNs~(see \citealt{Ricci2023} and references therein), which exhibit dramatic variability on the timescale of years to decades, likely driven by extreme changes in the accretion rate of the disk~\citep[e.g.,][]{Ruan2016, Noda2018, graham2020, Panda2024b}.
Such systems underscore the limitations of assuming stationarity in modeling AGN variability.
Future development of \pkg~will explore non-stationary GP kernels, in which the kernel parameters evolve with time rather than remaining fixed~\citep[e.g.,][]{Noack2021}.

\section{Summary \& Conclusion}\label{sec:conclusion}
We present \pkg~---a new software package designed to leverage multiband AGN light curves for joint inference of intrinsic stochastic variability properties and interband continuum lags.
We evaluate the performance of \pkg~using both simulated LSST AGN light curves sampled at the cadence of the WFD survey, \edit{and multiwavelength light curves of real AGNs}.
\edit{The \pkg~software is available on GitHub\footnote{\texttt{EzTaoX} codebase: \url{https://github.com/ywx649999311/EzTaoX}.} under a MIT License and version 0.1 is archived in Zenodo \citep{eztaox}.}
The key advancements and capabilities of \pkg~are summarized as follows:

\begin{itemize}
    \item \pkg~is $\sim10^2-10^3$ times faster than \jl~for LSST WFD light curves, and more than $\sim10^4\times$ faster for DDF light curves (see Figure~\ref{fig:bm} and Section~\ref{subsec:scalability}).
    
    \item \pkg~and \jl~show comparable reliability in recovering simulated interband lags. Specifically, lags can be recovered to within $\sim\,$2 days ($1\sigma$) using the first 3 years of LSST WFD observations (assuming a fiducial \taudrw~of 100 days, a \sigmadrw~of 0.112 mag, and a $g$-band mean magnitude of 21), and to within $\sim\,$0.7 day ($1\sigma$) with the full 10-year dataset (see Figure~\ref{fig:lag_recover}).

    
    \item \pkg~provides more reliable recovery of the simulated \taudrw~and \sigmadrw~than \jl. In particular, \jl~produces systematically biased estimates, whereas \pkg~recovers unbiased results across the tested range~(see Figure~\ref{fig:tau_recover} and Figure~\ref{fig:amp_recover}).

    \item \pkg's multiband modeling capability enables robust recovery of \taudrw~down to hourly timescales with simulated LSST WFD light curves, outperforming single-band modeling. This precision enhances the ability to identify low-mass AGNs (e.g., those in dwarf galaxies) through variability (see Figure~\ref{fig:tau_recover_hr}).
    
    \item \pkg~supports a broad class of GP kernels beyond the DRW model for joint modeling of AGN stochastic variability and interband lags, offering greater modeling flexibility (see Figure~\ref{fig:dho_demo} for an example). 
    
\end{itemize}

In addition to characterizing AGN stochastic variability and measuring continuum lags, \pkg~is applicable to a variety of other time-domain astrophysical studies.
For example, \pkg~can be applied to measure BLR emission line lags from spectroscopic RM light curves~\citep{blandford1982}, or to detect negative lags that may arise from inward propagation of accretion rate fluctuations~\citep{Secunda2023}---both of which typically occur on timescales much longer than the continuum lags.
Moreover, \pkg~can be used to measure time delays between the multiple images of gravitationally lensed quasars, providing an important avenue for constraining the Hubble constant~\citep{Blandford1992}.
Finally, interband lags have been shown to correlate with AGN continuum luminosity~\citep[e.g.,][]{sergeev2005, netzer2021, guo2022, Panda2024a}; once properly calibrated, this lag-luminosity relation may enable the estimation of AGN physical properties using photometric time-domain data alone.

\vspace{1cm}
\edit{We thank the anonymous referee for the thoughtful comments that helped us improve the paper.}
We thank Maurizio Paolillo \edit{and Thaisa Storchi Bergmann }for providing comments to an early draft of this manuscript. 
W.Y. acknowledges support from the Dunlap Institute for Astronomy \& Astrophysics at the University of Toronto. 
J.J.R.\ acknowledges support from the Canada Research Chairs (CRC) program, the NSERC Discovery Grant program, the Canada Foundation for Innovation (CFI), and the Qu\'{e}bec Minist\`{e}re de l'\'{E}conomie et de l'Innovation.
C.J.B. is supported by an NSF Astronomy and Astrophysics Postdoctoral Fellowship under award AST-2303803. This material is based upon work supported by the National Science Foundation under Award No. 2303803. This research award to NSF is partially funded by a generous gift of Charles Simonyi to the NSF Division of Astronomical Sciences. The award is made in recognition of significant contributions to Rubin Observatory’s Legacy Survey of Space and Time.
R.J.A., F.E.B, and S.S.S. gratefully acknowledge support from ANID: CATA BASAL FB210003 (R.J.A., F.E.B., S.S.S.); Millennium Science Initiative AIM23-0001 (F.E.B.); and FONDECYT Regular 1231718 (R.J.A.) and 1241005 (F.E.B.).
D.D. acknowledges PON R\&I 2021, CUP E65F21002880003, and Fondi di Ricerca di Ateneo (FRA), linea C, progetto TORNADO.
L.H.G. acknowledges financial support from ANID programs: FONDECYT Iniciaci\'on 11241477, Millennium Science Initiative Programs NCN2023-002, and AIM23-0001.
D.I. and A.B.K. acknowledge funding provided by the University of Belgrade - Faculty of Mathematics (the contract 451-03-136/2025-03/200104) through the grant by the Ministry of Science, Technological Development and Innovation of the Republic of Serbia.
S.P. is supported by the international Gemini Observatory, a program of NSF NOIRLab, which is managed by the Association of Universities for Research in Astronomy (AURA) under a cooperative agreement with the U.S. National Science Foundation, on behalf of the Gemini partnership of Argentina, Brazil, Canada, Chile, the Republic of Korea, and the United States of America.

Based on observations obtained with the Samuel Oschin Telescope 48-inch and the 60-inch Telescope at the Palomar Observatory as part of the Zwicky Transient Facility project. ZTF is supported by the National Science Foundation under Grants No. AST-1440341 and AST-2034437 and a collaboration including current partners Caltech, IPAC, the Oskar Klein Center at Stockholm University, the University of Maryland, University of California, Berkeley , the University of Wisconsin at Milwaukee, University of Warwick, Ruhr University, Cornell University, Northwestern University and Drexel University. Operations are conducted by COO, IPAC, and UW.

\software{
        pandas \citep{pandas2010},
        numpy \citep{numpy2020},    
        scipy \citep{scipy2020},
        matplotlib \citep{matplotlib2007},
        astropy \citep{astropy2013, theastropycollaboration2018, astropycollaboration2022},  
        emcee \citep{foreman-mackey2013},
        eztao \citep{yu2022},
        tinygp \citep{foreman-mackey2024},
        numpyro \citep{bingham2019pyro, phan2019composable}
        eztaox \citep{eztaox} 
         }

\bibliography{My_Library, lag_ref}{}
\bibliographystyle{aasjournal}

\appendix
\section{Additional \pkg~Experiments}\label{appendix:more_eztao_tests}
\begin{figure}
    \centering
    \includegraphics[width=0.5\linewidth]{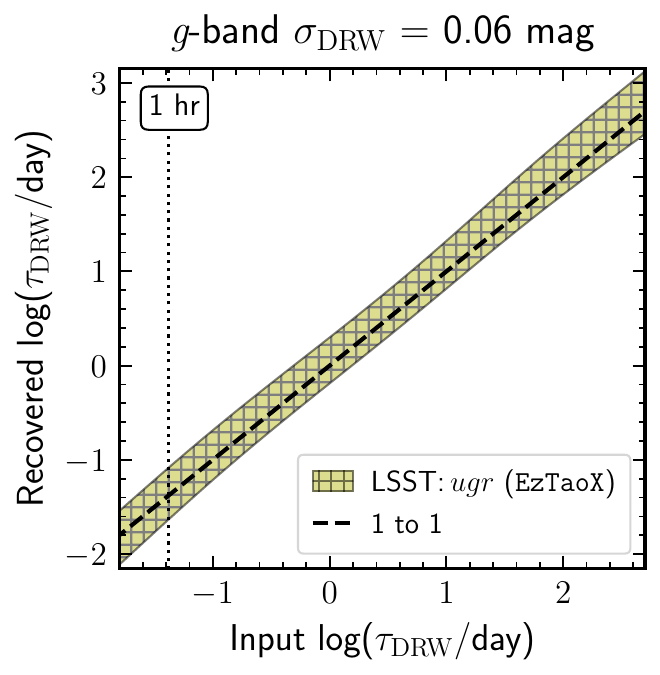}
    \caption{Recovered versus input \taudrw~from multiband ($u$, $g$, and $r$) light curves with \pkg. The 3 year LSST WFD $g$-band light curves are simulated with \taudrw~$=100$ days and \sigmadrw~$=$ 0.06 mag (half the amplitude of the light curves in Figure~\ref{fig:tau_recover}).
    Even when the intrinsic \sigmadrw~is only a few times the photometric uncertainty, \pkg~robustly recovers the input \taudrw~down to one hour.}
    \label{fig:a1_tau}
\end{figure}

\begin{figure}
    \centering
    \includegraphics[width=0.5\linewidth]{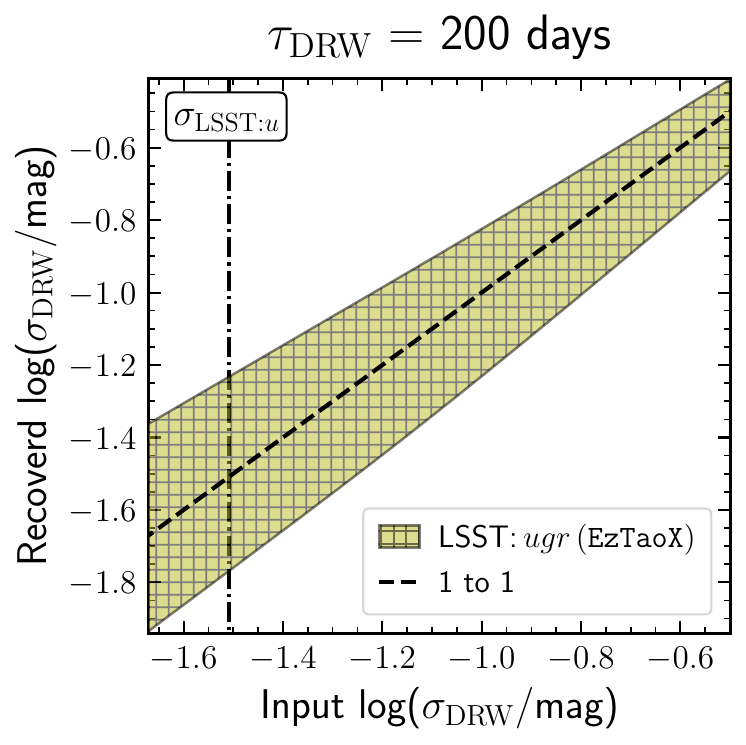}
    \caption{Recovered versus input \sigmadrw~from multiband ($u$, $g$, and $r$) light curves with \pkg. The 3 year LSST WFD $g$-band light curves are simulated with \taudrw~$=200$ days and \sigmadrw~$=$ 0.112 mag. Consistent with Figure~\ref{fig:amp_recover}, \pkg~yields unbiased estimates of \sigmadrw~across the full simulated range.}
    \label{fig:a1_amp}
\end{figure}

We perform additional experiments to demonstrate the robustness of \pkg~in recovering DRW parameter under different regimes. 
To test its performance in recovering \taudrw~at lower intrinsic variability amplitudes, we simulate a set of light curves similar to those in Section~\ref{subsubsec:drw_tau_recover}, but with the $g$-band \sigmadrw~reduced to 0.06 mag---half the value adopted in Section~\ref{subsubsec:drw_tau_recover} and only about four times the average photometric uncertainty. As shown in Figure~\ref{fig:a1_tau}, \pkg~still robustly recovers \taudrw~down to one hour with 3 years of LSST WFD light curves.
To evaluate its performance in recovering \sigmadrw~given larger intrinsic \taudrw, we simulate another set of light curves similar to those in Section~\ref{subsubsec:drw_amp_recover}, but with the $g$-band \taudrw~increased to 200 days---twice the value adopted in Section~\ref{subsubsec:drw_amp_recover}. As shown in Figure~\ref{fig:a1_amp}, \pkg~continues to provide unbiased estimates of \sigmadrw~across the full simulated range.

\section{Additional \jl~Experiments}\label{appendix:more_jl_tests}

\begin{figure}
    \centering
    \includegraphics[width=0.5\linewidth]{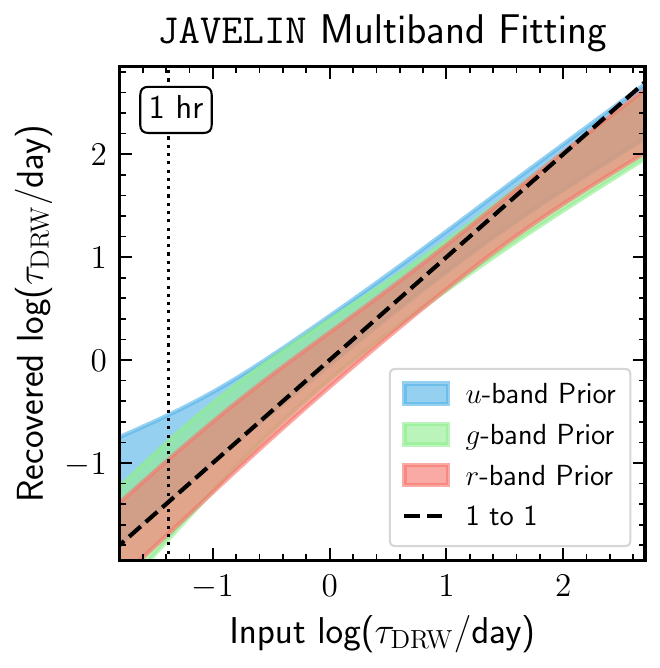}
    \caption{Recovered versus input \taudrw~from multiband ($u$, $g$, and $r$) light curves with \jl. The blue, green, and red regions show the central 68\% intervals for the distributions of the recovered \taudrw~with the priors derived from $u$-, $g$-, and $r$-band light curves, respectively. The recovered \taudrw~distributions differ depending on the band used to generate the priors.}
    \label{fig:a2_tau}
\end{figure}

\begin{figure}
    \centering
    \includegraphics[width=0.5\linewidth]{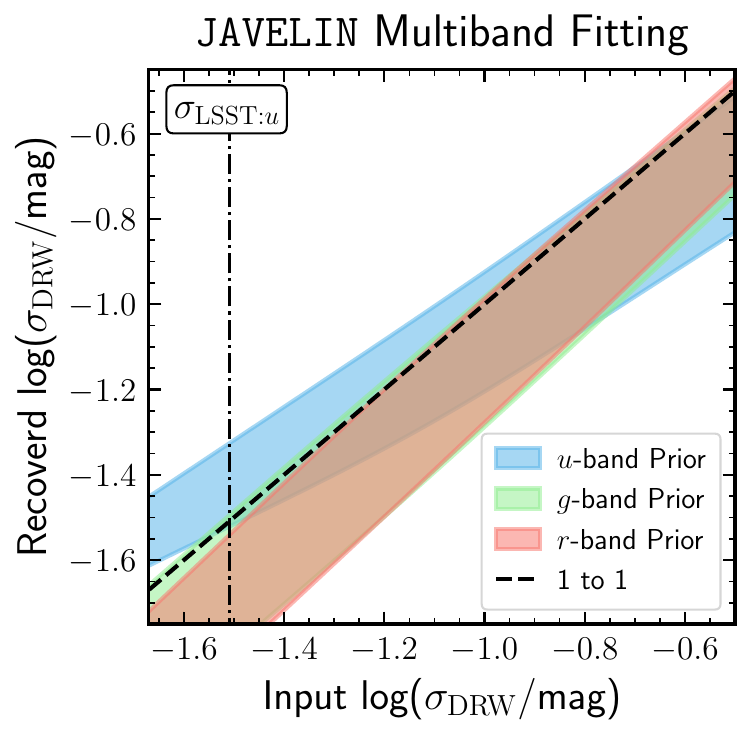}
    \caption{Recovered versus input \sigmadrw~from multiband ($u$, $g$, and $r$) light curves with \jl. The blue, green, and red regions show the central 68\% intervals for the distributions of the recovered \sigmadrw~with the priors derived from $u$-, $g$-, and $r$-band light curves, respectively. The recovered \sigmadrw~distributions differ depending on the band used to generate the priors.}
    \label{fig:a2_amp}
\end{figure}

The distribution of DRW parameters recovered by \jl's multiband fitting depends on the properties of the light curves (e.g., cadence and mean photometric uncertainty) used in its initial fitting step, where a prior distribution of \taudrw~and \sigmadrw~is generated from a selected single-band light curve. 
To illustrate this sensitivity, we refit the light curves from Section~\ref{subsubsec:drw_tau_recover} and Section~\ref{subsubsec:drw_amp_recover}, generating the prior distributions from the $u$- and $g$-band light curves. 
Figures~\ref{fig:a2_tau} and \ref{fig:a2_amp} show the resulting distributions of \taudrw~and \sigmadrw~recoveries, highlighting that the final DRW parameter distributions differ depending on the band used to generate the priors.
Moreover, more than 40\% of \jl~fits fail the cleaning cuts when the $u$-band is used to generate the priors. This likely arises because the $u$-band has the poorest temporal sampling and highest photometric uncertainty, making the priors from \jl's first fitting step unreliable.

\end{CJK*}
\end{document}